\documentclass[lettersize,journal]{IEEEtran}
\usepackage{amsmath,amsfonts}
\usepackage{algorithm}
\usepackage[noend]{algpseudocode}  % algorithmicx 宏包的一部分

\usepackage{array}
\usepackage{textcomp}
\usepackage{stfloats}
\usepackage{url}
\usepackage{verbatim}
\usepackage{graphicx}
\usepackage{cite}
\usepackage{booktabs}
\usepackage{multirow}
\usepackage{tabularx}
\usepackage{subcaption}
\usepackage{listings}
\begin{document}

\title{FLSSM: A  Federated Learning Storage Security Model with Homomorphic Encryption}
%基于同态加密的 Federated Learning 存储安全模型(
%\author{IEEE Publication Technology,~\IEEEmembership{Staff,~IEEE,}
	\author{Yang~Li,
		Chunhe~Xia,
		Chang~Li,
		Xiaojian Li
		and~Tianbo~Wang,~\IEEEmembership{Member,~IEEE,}% <-this % stops a space
		\thanks{Yang Li is with the School of Computer Science and Engineering, Beihang University, Beijing 100191, China (e-mail: johnli@buaa.edu.cn).}% <-this % stops a space
		\thanks{Chunhe Xia is with the Key Laboratory of Beijing Network Technology, Beihang University, Beijing 100191, China, and also with the Guangxi 	Collaborative Innovation Center of Multi-Source Information Integration and Intelligent Processing, Guangxi Normal University, Guilin 541004, China. (e-mail: xch@buaa.edu.cn).}% <-this % stops a space
		\thanks{Chang Li is with the School of Computer Science and Technology, Zhengzhou University of Light Industry, Zhengzhou 450000, China (e-mail: 3031169424@qq.com).}
		\thanks{Xiaojian Li is with the College of Computer Science and Information	Technology, Guangxi Normal University, Guilin 541004, China.}
		\thanks{Tianbo Wang is with the School of Cyber Science and Technology, Beihang University, Beijing 100191, China, and also with the Shanghai Key Laboratory 	of Computer Software Evaluating and Testing, Shanghai 201112, China} 
%		\thanks{Tianbo Wang is with the School of Cyber Science and Technology, Beihang University, Beijing 100191, China, and also with the Shanghai Key Laboratory 	of Computer Software Evaluating and Testing, Shanghai 201112, China} 
        % <-this % stops a space
\thanks{This paper was produced by the IEEE Publication Technology Group. They are in Piscataway, NJ.}% <-this % stops a space
\thanks{Manuscript received April 19, 2021; revised August 16, 2021.}}

\markboth{Journal of \LaTeX\ Class Files,~Vol.~14, No.~8, August~2021}%
{Shell \MakeLowercase{\textit{et al.}}: A Sample Article Using IEEEtran.cls for IEEE Journals}


\maketitle

\begin{abstract}
%This document describes the most common article elements and how to use the IEEEtran class with \LaTeX \ to produce files that are suitable for submission to the IEEE.  IEEEtran can produce conference, journal, and technical note (correspondence) papers with a suitable choice of class options. 
%基于同态加密的联邦学习因其具有较高的安全性，可以更好的保护用户数据隐私而受到广泛关注。然而，其加密计算的特点导致了三个棘手的问题："computation-efficiency","attack-tracing" and "contribution-assessment". 第一个指的是模型聚合时加密计算的效率，第二个指的是加密状态下对恶意攻击的追溯，第三个指的是加密后本地模型的贡献评估的公平性。本文提出了a federated learning storage security model with homomorphic encryption (FLSSM) 保护联邦学习模型隐私，以解决上述三个问题。首先，我们利用不同的节点并行聚合本地模型，以提高聚合效率；其次，引入可信的监管节点在全局模型受到攻击时对本地模型进行审查，实现对恶意攻击的追溯。最后，基于可信的训练时间对本地训练节点进行奖励。Experiments on multiple real-world datasets show that 我们的模型在xxx和xxx指标上都明显优于最先进的基线模型。

Federated learning based on homomorphic encryption has received widespread attention due to its high security and enhanced protection of user data privacy. However, the characteristics of encrypted computation lead to three challenging problems: ``computation-efficiency", ``attack-tracing" and ``contribution-assessment". The first refers to the efficiency of encrypted computation during model aggregation, the second refers to tracing malicious attacks in an encrypted state, and the third refers to the fairness of contribution assessment for local models after encryption.
This paper proposes a federated learning storage security model with homomorphic encryption (FLSSM) to protect federated learning model privacy and address the three issues mentioned above. First, we utilize different nodes to aggregate local models in parallel, thereby improving encrypted models' aggregation efficiency. Second, we introduce trusted supervise nodes to examine local models when the global model is attacked, enabling the tracing of malicious attacks under homomorphic encryption. Finally, we fairly reward local training nodes with encrypted local models based on trusted training time.
Experiments on multiple real-world datasets show that our model significantly outperforms  baseline models in terms of both efficiency and security metrics.
\end{abstract}


%1. 利用同态加密 将模型分层存储在不同节点 不同的节点分别对不同层的参数进行同态加密 提高效率
%2. 聚合节点在同态加密的情况下 将不同层已聚合好的模型合并在一起 形成新的全局模型
%
%- 3层结构：聚合节点 --> 边缘聚合节点 --> 本地训练节点
%- 节点：1. 聚合节点 2. 边缘聚合节点 3. 本地训练节点 4. 时间戳服务器
%- 本地模型参数按照不同的层分别存储在边缘计算节点中
%- 基于门限加密的同态加密（解决历史模型访问问题）：
%```五、实施步骤示例（以门限加密为例）
%初始化：
%- 客户端生成HE密钥对，将私钥拆分为多个分片（如5个分片需3个协作解密）。
%- 分片分发给多个可信节点（如监管机构、客户端代表）。
%训练阶段：
%客户端使用HE公钥加密本地模型参数，上传至聚合节点。
%聚合节点执行密文聚合，生成全局模型密文。
%历史访问阶段：
%- 聚合节点向至少3个分片持有者发送解密请求。
%- 分片持有者联合计算解密全局模型密文，返回明文给聚合节点。
%```
%```
%用户操作：                                验证者操作：
%1. 生成文件 →                             1. 获取原始文件、签名、公钥、时间戳（可选）。
%2. 计算文件哈希 →                         2. 重新计算文件哈希。
%3. 用私钥加密哈希（生成签名）→            3. 用公钥解密签名，得到签名时的哈希。
%4. 向 TSA 申请时间戳（可选）→   （RFC 3161协议 需改进 时间戳对 将一对时间戳作为一个整体返回请求发起节点 保证了开始时的文件是上一轮的全局模型 结束的文件是更新后的本地模型 时间戳对可以直接表示训练所需的时间）         4. 验证时间戳合法性（若存在）。
%4.1 由于共识算法根据时间间隔选择聚合节点 因此节点会尽可能让时间间隔变小 为了防止节点推迟发送全局模型的hash 因此本地训练节点的开始时间戳从收到全局模型开始计算 由聚合节点/区块链/可信节点 或其他的节点发送给时间戳服务器（访问数据库/存储节点/Storj/IPFS/接收区块 的时间）(利用智能合约记录每个节点获取全局模型的时间)
%4.2 而节点会尽可能提前更新后的本地模型的时间戳 对更新后本地模型的hash验证可以保证节点是在本地模型训练完成后申请的时间戳
%4.3 两段式时间戳协议 第一阶段收到时间戳请求作为开始 第二阶段收到时间戳请求作为结束
%4.4 结构：开始hash|开始时间戳|结束hash|结束时间戳
%5. 将文件、签名、时间戳发送给验证者。    5. 对比当前哈希与签名哈希是否一致。
%→ 一致：文件未篡改，签名有效。
%→ 不一致：文件已被修改。
%```
%
%创新点：1、基于同态加密的分布式聚合模型 2、根据时间间隔的共识算法 3、基于门限加密算法的模型访问控制

%\section{ }

%\subsection{ Abstract}
%\subsubsection{ Keywords}
%Federated Learning, Blockchain, Homomorphic Encryption,门限加密
\section{ Introduction}
%
%联邦学习（Federated Learning, FL）作为一种新兴的机器学习范式，旨在保护用户数据的隐私安全。在传统的集中式机器学习中，用户数据需要上传到中央服务器进行模型训练，这不仅增加了数据泄露的风险，还可能导致用户对数据隐私的担忧。联邦学习中用户只需共享模型参数，而将数据留在本地，从而有效地避免了数据的直接传输和泄露，因此受到越来越多的关注。然而，联邦学习是一把双刃剑。从积极的方面来看，联邦学习将数据保留在本地设备上的方式大大减少了数据在传输过程中的暴露风险，打消了人们关于数据隐私安全的担忧，提高了人们参与模型训练的积极性。从消极的方面来看，恶意攻击者仍可能通过中间人攻击、推测攻击等方式窃取模型参数后推断用户的数据分布或特定数据是否存在，从而造成隐私泄露，加大人们对于数据安全的恐慌，因此借由加密技术保护本地模型参数隐私的方法变得十分重要。
%
%其中，广泛使用的加密技术可分为同态加密、安全多方加密等，两者均具有较高的安全性，但同时具有较高的计算复杂度，需要消耗大量计算资源。安全多方计算技术需要多个参与方合作，且通信开销较高。同态加密允许单个参与方在本地对加密数据进行计算，而不需要与其他参与方进行交互，减少了对其他参与方的依赖。但基于同态加密的联邦学习中存在3个不可避免的问题：加密模型聚合效率问题，加密模型访问控制问题和加密模型贡献评估问题。由于同态加密技术对本地模型进行了加密，聚合服务器需要对本地模型进行同态计算，这对聚合服务器的计算能力提出了较高要求，当存在大量本地模型时，聚合服务器的计算能力和资源的需求会显著增加。由于本地模型在加密后发送给聚合服务器，聚合服务器在聚合时进行同态计算，无法对加密后的本地模型进行验证，若存在恶意攻击者实施模型投毒攻击或后门攻击，则模型性能可能显著下降，使得全局模型面临安全威胁。而针对用户在联邦学习中的贡献进行评估对于联邦学习的可持续发展具有重要意义，公平的激励机制可以促进用户积极参与联邦学习的模型训练，但基于同态加密的联邦学习中用户共享的是加密后的本地模型，现有方法如基于梯度、基于性能、基于数据质量等方法无法满足需求，因此如何在不泄露用户数据隐私的前提下对用户贡献进行公平评估成为面临的问题之一。
%
%
%为了解决上述问题，我们提出了A  Federated Learning Storage Security Model with Homomorphic Encryption (FLSSM),该模型由3个主要部分组成:基于同态加密的分布式聚合机制(HAM)，基于Shamir秘密共享的模型访问控制机制(MACM)和基于可信时间间隔的激励机制(IMTTI)。HAM中由不同的边缘节点并行聚合本地模型的不同层，提高了同态加密状态下联邦学习的聚合效率。MACM中允许监管节点在一定条件下对本地模型进行审查，从而防止加密状态下本地模型的恶意攻击。IMTTI中利用可信的本地模型训练时长对用户进行奖励，在不泄露模型隐私的前提下提出了一种新的可信模型贡献评估方法。

Federated Learning (FL) \cite{Kairouz2019AdvancesAO,Li2019FederatedLC} has emerged as a novel machine learning paradigm designed to protect the privacy and security of user data \cite{Mothukuri2021ASO}. In traditional centralized machine learning, user data must be uploaded to a central server for model training, which not only increases the risk of data breaches but may also raise user concerns about data privacy \cite{Neal2006PatternRA}. In federated learning, users only need to share model parameters while keeping their data local, effectively avoiding direct data transmission and leakage, thus attracting increasing attention \cite{zhang2021survey}.

However, federated learning is a double-edged sword. On the positive side, keeping data on local devices significantly reduces the risk of exposure during transmission, alleviating concerns about data privacy security and increasing people's willingness to participate in model training \cite{kairouz2021advances}. On the negative side, malicious attackers may still steal model parameters through man-in-the-middle attacks, inference attacks, and other methods to deduce users' data distribution or determine whether specific data exists, causing privacy leakage and increasing fears about data security \cite{wen2023survey}. Therefore, methods that use encryption technologies to protect local model parameter privacy have become increasingly important.

Widely used encryption technologies in federated learning can be divided into homomorphic encryption, secure multi-party computation, and others \cite{fang2021privacy}. Both provide high security but come with high computational complexity, requiring substantial computational resources. Secure multi-party computation  requires cooperation between multiple parties and has high communication overhead \cite{mugunthan2019smpai}. Homomorphic encryption allows a single participant to compute on encrypted data locally without interacting with other parties, reducing dependency on other participants \cite{zhang2020batchcrypt}. However, federated learning based on homomorphic encryption faces three unavoidable issues: encrypted model aggregation efficiency, encrypted model access control, and encrypted model contribution assessment \cite{liu2022privacy}.

Since homomorphic encryption encrypts local models, the aggregation server needs to perform homomorphic computations on these models, placing high demands on the server's computational capabilities \cite{xie2024efficiency}. When numerous local models exist, the requirements for the aggregation server's computational power and resources increase significantly. Since local models are encrypted before being sent to the aggregation server, which performs homomorphic computations during aggregation, the server cannot verify the encrypted local models. If malicious attackers implement model poisoning or backdoor attacks, model performance may significantly decrease, posing security threats to the global model. Assessing user contributions in federated learning is crucial for sustainable development, as fair incentive mechanisms can encourage active user participation in model training. However, in federated learning with homomorphic encryption, users share encrypted local models, making existing methods based on gradients, performance, or data quality inadequate \cite{rieyan2024advanced}. Therefore, how to fairly evaluate user contributions without compromising data privacy becomes one of the challenges.

To address these issues, we propose A Federated Learning Storage Security Model with Homomorphic Encryption (FLSSM), consisting of three main components: a Homomorphic-encryption-based Aggregation Mechanism (HAM), a Model Access Control Mechanism based on Shamir's Secret Sharing (MACM), and an Incentive Mechanism based on Trusted Time Intervals (IMTTI). In HAM, different edge nodes parallelly aggregate different slices of local models, improving aggregation efficiency under homomorphic encryption. MACM allows regulatory nodes to review local models under certain conditions, preventing malicious attacks on local models in encrypted states. IMTTI rewards users based on trusted local model training duration, proposing a new trusted model contribution assessment method without leaking model privacy.
Our contributions can be summarized as follows.

\begin{enumerate}
%	\item We propose the FLSSM model to address the 基于同态加密的联邦学习中聚合效率、恶意攻击和奖励分配的问题。Afterward, we design a 聚合算法，a 访问控制机制和 a 激励机制来分别解决他们。
%	\item We 提出了一个新的聚合算法，通过并行聚合提高了同态加密下联邦学习聚合效率，同时设置可信节点对本地模型进行审查，对恶意攻击进行追溯。
%	\item We 提出了一个新的激励机制，通过模型训练时间对模型贡献进行可信评估，在不泄露模型隐私的前提下实现了用户的公平激励。
%	\item Experiments on xxx个 real-world datasets show that our model outperforms state-of-the-art baseline works in both accuracy, loss and F1 metrics.

\item We propose the FLSSM model to address the issues of aggregation efficiency, malicious attacks, and reward allocation in federated learning based on homomorphic encryption. Afterward, we design an aggregation algorithm, an access control mechanism, and an incentive mechanism to solve these problems respectively.
\item We present a novel aggregation algorithm that improves the aggregation efficiency of federated learning under homomorphic encryption through parallel aggregation, while establishing trusted nodes to review local models and trace malicious attacks.
\item We introduce a new incentive mechanism that conducts reliable assessment of model contributions through model training time, achieving fair user incentivization without compromising model privacy.
\item Experiments on two real-world datasets demonstrate that our model's effectiveness.
\end{enumerate}

The rest of this article is organized as follows. We first give a comprehensive review of related works in Section \ref{Related-Works}. Next, we give the preliminaries for this article in Section \ref{Preliminaries}, then formalize the problems and present technical details of FLSSM in Section \ref{Methodology}. After that, we conduct a series of experiments on four public datasets to evaluate FLSSM in Section \ref{Performance-Evaluation}. Finally, Section \ref{Conclusion} concludes this work.
%我们的主要贡献
%文章主要结构
%场景图（要解决的问题及这个模型适用的场景 要在什么场景下解决这个问题）
\section{ Related Works}
\label{Related-Works}
%针对三个问题的相关工作 
%1. 针对加密效率问题的工作 分为哪几种类型 有优缺点
%2. 针对加密模型访问控制问题的工作 分为哪几种类型 有优缺点
%3. 针对加密模型贡献评估问题的工作 分为哪几种类型 有优缺点
%\subsection{加密模型聚合效率}
%
%基于加密技术保护联邦学习模型隐私可以提高安全性，但同样需要进行大量计算，这也降低了联邦学习的聚合效率。
%为了提高加密后联邦学习的效率，
%Chengliang Zhang et al. \cite{c783cdc03a32e5094affa7eef710459aac599aaf}将一批 量化的梯度编码为一个长整数并一次性加密，并开发了新的量化和编码方案以及一种新的梯度裁剪技术.Ren-Yi Huang \cite{9e889c1c151d03e9d6db6d877f471f0c9756cca0}引入关键模型参数的选择性加密，以减少计算/通信开销，解决 HE 联合学习的效率问题。Weizhao Jin et al. \cite{bibid}提出有选择地加密敏感参数，以减少计算和通信开销，为大模型实现可扩展的基于同态加密的聚合。Sean Choi et al. \cite{10b7e45e27b440aacc0a1d934ff5f21c0ed8e8f6} 通过利用SmartNIC卸载计算密集型同态加密任务、降低CPU使用率和改善资源分配来提高效率。Kai Cheng et al. \cite{ebd92551aa8937327be29597aa9af3832812c7fc} 利用英特尔QAT硬件加速器、错误反馈梯度压缩和霍夫曼编码来简化加密和聚合过程。Valentino Peluso et al. \cite{3622d828324b053f76d4bc547a98fc8210eab585}引入了私有张量冻结（PTF）门控方案，可以随着时间的推移降低加密/解密、通信和服务器聚合的复杂性。但现有方法在模型训练过程中的隐私性和聚合效率难以达到平衡，通常为了提高效率需要牺牲一定的隐私性。
\subsection{Encrypted Model Aggregation Efficiency}
Protecting federated learning model privacy with encryption technologies can enhance security but also requires extensive computation, which reduces the aggregation efficiency of federated learning.
To improve the efficiency of federated learning after encryption,
Chengliang Zhang et al. \cite{c783cdc03a32e5094affa7eef710459aac599aaf} encoded a batch of quantized gradients into a single long integer for one-time encryption, and developed new quantization and encoding schemes as well as a new gradient clipping technique. Ren-Yi Huang \cite{9e889c1c151d03e9d6db6d877f471f0c9756cca0} introduced selective encryption of key model parameters to reduce computational/communication overhead, addressing the efficiency issues in HE-based federated learning.  Sean Choi et al. \cite{10b7e45e27b440aacc0a1d934ff5f21c0ed8e8f6} improved efficiency by offloading computationally intensive homomorphic encryption tasks to SmartNICs, reducing CPU utilization and improving resource allocation. Kai Cheng et al. \cite{ebd92551aa8937327be29597aa9af3832812c7fc} utilized Intel QAT hardware accelerators, error-feedback gradient compression, and Huffman coding to simplify encryption and aggregation processes. Valentino Peluso et al. \cite{3622d828324b053f76d4bc547a98fc8210eab585} introduced a Private Tensor Freezing (PTF) gating scheme that reduces the complexity of encryption/decryption, communication, and server aggregation over time. However, existing methods struggle to balance privacy and aggregation efficiency during model training, often sacrificing some privacy to improve efficiency.
\subsection{Encrypted Model Access Control}
Encrypted model parameters cannot be accessed, which creates security vulnerabilities during federated learning model aggregation, as malicious attackers may poison local models before sending them to the aggregation server. Existing model access control mechanisms can be categorized into multi-key homomorphic encryption and authentication/authorization approaches. Yuxuan Cai et al. \cite{Cai2024SecFedAS} introduced "EMK-BFV," combining multi-key homomorphic encryption with Trusted Execution Environments (TEEs) to enhance privacy, access control, and efficiency in federated learning. Jiachen Shen \cite{50f159cb5e5b42dd94a125e30b9bb3c07dc13e2c} expanded the scope of addressed privacy risks by encrypting with an aggregated public key and requiring joint decryption among participants. Xueyin Yang \cite{735a17688e68ea922f193f835d61b4120c1737c7} designed a multi-private-key secure aggregation algorithm implementing homomorphic addition operations, allowing servers and clients to freely choose public-private key pairs, making it more suitable for deep models. Additionally, they compressed multi-dimensional data into one dimension, significantly reducing encryption/decryption time and ciphertext transmission communication.
Caimei Wang \cite{953bdb701c020617028e6a64675f8c84595a653d} proposed a fingerprint-based subkey verification algorithm (FKM) to generate unique fingerprints for each subkey, while designing a gradient protection scheme to achieve higher security levels and reduce encryption overhead. Chun-I Fan et al. \cite{10.1145/3540199} proposed an identity-based multi-receiver homomorphic proxy re-encryption (IMHPRE) scheme that utilizes homomorphic addition and re-encryption to provide improved encrypted data processing and access control. When adopting this scheme, participants can encrypt directly using public identities.
Jing Wang et al. \cite{10566968} proposed a hierarchical cloud-edge orchestration federated learning architecture for IoT, designing an IoT knowledge sharing method based on multi-level access control encryption to ensure knowledge confidentiality. Hui Lin et al. \cite{9538823} proposed an attribute-based secure access control mechanism, discovering the relationship between users' social attributes and their trustworthiness, where users' trustworthiness depends on their social influence, which is then transformed into trust levels. They used federated deep learning to obtain optimal thresholds for trust levels and related access control parameters to improve access control accuracy and enhance privacy protection.

While existing federated learning model access control mechanisms improve privacy protection and security to some extent, they often involve complex key management. A key challenge is how to enhance the flexibility of control mechanisms to accommodate the dynamic nature of nodes in federated learning.

\subsection{Encrypted Model Contribution Assessment}
As local models are sent to the aggregation server in an encrypted state, it is difficult to assess the contributions of local models based on existing methods such as Shapley values or accuracy. Liangjiang Chen et al. \cite{6faeb87c2e8d7ede87d954a5a18f772bad1617c8} addressed encryption efficiency and contribution assessment issues by asynchronously allocating gradient weights securely based on user data quality. Ruizhe Yang et al. \cite{10475694} proposed a federated learning method combining blockchain, homomorphic encryption, and reputation. Edge nodes with local data can train encrypted models using homomorphic encryption, and their contributions to aggregation are assessed through a reputation mechanism. Both models and reputations are recorded and verified on the blockchain through a consensus process, with rewards determined according to incentive mechanisms.
Longyi Liu et al. \cite{10831241} integrated federated learning into blockchain consensus protocols, proposing an FRConsensus algorithm based on model evaluation and stake election to overcome problems of passive participation and resource waste during model training. Additionally, they introduced model watermarking and ECC public key encryption mechanisms to protect parameter transmission. Biwen Chen et al. \cite{9780227} first constructed an efficient non-interactive designated decryptor functional encryption scheme that protects training data privacy while maintaining high communication performance. Then, by combining this framework with a carefully designed blockchain, they proposed a blockchain-based federated learning framework providing fair compensation for medical image detection.
Guilin Guan et al. \cite{10838122} encrypted weights using multi-key homomorphic encryption to resist data recovery attacks launched by malicious edge servers and devices. Meanwhile, based on marginal loss techniques, they detected malicious clients or those uploading low-quality contributions, and assessed edge devices' contribution levels based on Shapley techniques to distribute rewards to participants. Ke Geng et al. \cite{10476158} utilized arithmetic sharing to achieve submodel reconstruction and utility evaluation required in gradient Shapley under privacy protection, using shuffling and asymmetric encryption to ensure the privacy of test data collected from participating clients. Yingxin Li et al. \cite{10707464} proposed a personalized residual federated secure learning scheme (PRFSL) based on homomorphic encryption and edge computing to guarantee security, timeliness, integrity of task data, and privacy needs of group workers, thereby improving encryption efficiency. Finally, they proposed a personalized privacy incentive mechanism based on evolutionary game theory to improve overall service utility.

Existing research has made some progress in incentive mechanisms for encrypted models, but in practical applications, due to encryption and privacy protection constraints, it remains challenging to balance privacy protection and the verifiability of contribution assessment.
\section{ Preliminaries}
\label{Preliminaries}
%\subsection{ 同态加密}
%\label{sec:pre-homo}
%
%同态加密指的是加密后的密文满足同态运算性质的加密算法，即对数据进行加法或乘法运算后进行加密，和对加密后的数据进行加法或乘法运算，得到的结果是等价的。对于任意加密函数$\vartheta$，若对于任意数据A和数据B满足$Dec_{\vartheta}(Enc_{\vartheta}(A)\odot Enc_{\vartheta}(B)) = A \oplus B$,则认为 $\vartheta$具有同态性。在此基础上，同态加密根据支持的运算方式不同可分为加法同态和乘法同态，加法同态表示加密后的密文支持加法运算，乘法同态表示加密后的密文支持乘法运算。对于仅支持加法、仅支持乘法、或支持有限次加法或乘法运算的加密函数，为半同态加密或部分同态加密；而同时满足无限次加法或乘法运算的加密函数，为全同态加密。
%
%
%现有主流同态加密算法包括CKKS、BFV、Paillier算法，CKKS算法基于环上的学习问题（RLWE），支持近似同态加密，具有支持高效的浮点数运算的优点，但同时存在一定的误差积累问题，且密文扩展会导致计算效率下降。
%BFV算法基于环上的学习问题（RLWE），支持全同态加密，具有支持精确的整数运算的优点，但同时计算复杂度较高，密文扩展严重，导致存储和传输成本增加。
%Paillier算法基于大数分解问题，支持部分同态加密，具有实现简单，支持加法同态的优点，但同时仅支持加法同态，不支持乘法同态，限制了其应用场景。

\subsection{Homomorphic Encryption}
\label{sec:pre-homo}
Homomorphic encryption refers to encryption algorithms where the ciphertexts satisfy homomorphic operational properties, meaning that encrypting data after performing addition or multiplication operations is equivalent to performing addition or multiplication operations on already encrypted data \cite{acar2018survey}. For any encryption function $\vartheta$, if for any data A and data B, it satisfies $Dec_{\vartheta}(Enc_{\vartheta}(A)\odot Enc_{\vartheta}(B)) = A \oplus B$, then $\vartheta$ is considered to have homomorphism. Based on this, homomorphic encryption can be classified according to the supported operations into additive homomorphism and multiplicative homomorphism. Additive homomorphism indicates that encrypted ciphertexts support addition operations, while multiplicative homomorphism indicates that encrypted ciphertexts support multiplication operations. Encryption functions that only support addition, only support multiplication, or support a limited number of addition or multiplication operations are known as semi-homomorphic or partially homomorphic encryption; while encryption functions that satisfy unlimited addition or multiplication operations are fully homomorphic encryption.

Existing mainstream homomorphic encryption algorithms include CKKS, BFV, and Paillier algorithms. The CKKS algorithm is based on the Ring Learning With Errors problem (RLWE) and supports approximate homomorphic encryption. It has the advantage of supporting efficient floating-point operations, but also suffers from accumulated errors and reduced computational efficiency due to ciphertext expansion.
The BFV algorithm is also based on the Ring Learning With Errors problem (RLWE) and supports fully homomorphic encryption. It has the advantage of supporting precise integer operations, but also has higher computational complexity and significant ciphertext expansion, leading to increased storage and transmission costs.
The Paillier algorithm is based on the large number factorization problem and supports partially homomorphic encryption. It has the advantages of simple implementation and support for additive homomorphism, but only supports additive homomorphism and not multiplicative homomorphism, limiting its application scenarios.

\subsection{Shamir's Secret Sharing}
\label{sec:pre-shamir}
Shamir's Secret Sharing is a threshold scheme based on polynomial interpolation, allowing a secret \( S \) to be divided into \( n \) shares, where at least \( t \) shares are required to recover the secret \cite{beimel2011secret}.
The main steps are as follows:
\begin{enumerate}
	
\item \textbf{Initialization}.
Suppose the data to be encrypted is $s$, which will be divided into $n$ shares, with at least $t$ shares required for decryption. Therefore, $n$ is the number of participants, and $t$ is the threshold in Shamir's secret sharing. Randomly select a large prime number $p$ such that $(p>s)$.

\item \textbf{Encryption}.
Arbitrarily select t-1 random numbers $(x_1,\cdots,x_n)$, construct a polynomial $f(x)$ of degree $t-1$, where $f(0) = s$:
\begin{equation}
	f(x) = s + a_1 x + a_2 x^2 + \cdots + a_{t-1} x^{t-1} \pmod{p}
\end{equation}
Hence, $( a_1, a_2, \dots, a_{t-1} )$ are random coefficients $( 0 \leq a_i < p )$.
Arbitrarily select $n$ different numbers $(x_1,\cdots,x_n)\in \mathbb{z}^+$, and calculate the corresponding $y_i = f(x_i)=
s + a_1 x_i + a_2 x_i^2 + \cdots + a_{t-1} x_i^{t-1} \pmod{p}
$. Each $x_i$ and its corresponding $y_i$ form a secret share, and $(x_i, y_i)$ is distributed to the corresponding $n$ participants.

\item \textbf{Decryption}.
Collect at least $t$ secret shares $(x_i, y_i)$, and use Lagrange interpolation to calculate $s = f(x = 0)$. The Lagrange basis polynomials $L_i (x)$ and the calculation of $s$ are shown as follows:
\begin{equation}
	L_i(x) = \prod_{\substack{1 \leq j \leq t \\ j \neq i}} \frac{x - x_j}{x_i - x_j}
\end{equation}
\begin{equation}
	s = \sum_{i=1}^t y_i L_i(0) \pmod{p}
\end{equation}

\end{enumerate}
%
%\subsection{ 可信时间戳}
%
%可信时间戳在RFC 3161 中被定义, 本质是将用户的电子数据的hash值与权威的时间源绑定，在此基础上通过时间戳服务器进行签名，产生不可伪造的时间戳文件，从而证明某个文件或数据在特定时间之前已经存在，同时通过哈希值和数字签名，可信时间戳可以确保文件在生成时间戳后未被篡改。可信时间戳依赖于可信的时间戳权威机构（TSA），这些机构通常由政府或行业认可，具有较高的信誉度。任何人都可以使用TSA的公钥验证时间戳的有效性，确保过程的透明性和公正性。The main steps are as follows:
%\begin{enumerate}
%\item \textbf{生成摘要}.为要生成时间戳的数据进行hash计算，得到数据摘要:
%\begin{equation}
%h = H(D)
%\end{equation}
%\item \textbf{发起时间戳请求}.
%用户向Time Stamping Authority(TSA)发送请求，包含数据的哈希值$h$，生成时间戳$\sigma$：
%
%\begin{equation}
%\sigma = Sign_{TSA}(T)
%\end{equation}
%Hence, $T = (h, time)$ where $h$是数据的哈希值，$time$是当前时间。
%\item \textbf{返回时间戳}.
%TSA返回 $(T, \sigma)$ 给用户。
%\item \textbf{验证时间戳}.
%用户接收到TSA返回的时间戳 $(T, \sigma)$后，可使用TSA的公钥验证签名 $\sigma$是否真实。
%\begin{equation}
%\text{Verify}_{\text{TSA}}(T, \sigma) = \text{True}
%\end{equation}
%\end{enumerate}

\begin{figure*}[!htb]
	\centering
	\includegraphics[width=0.9\textwidth]{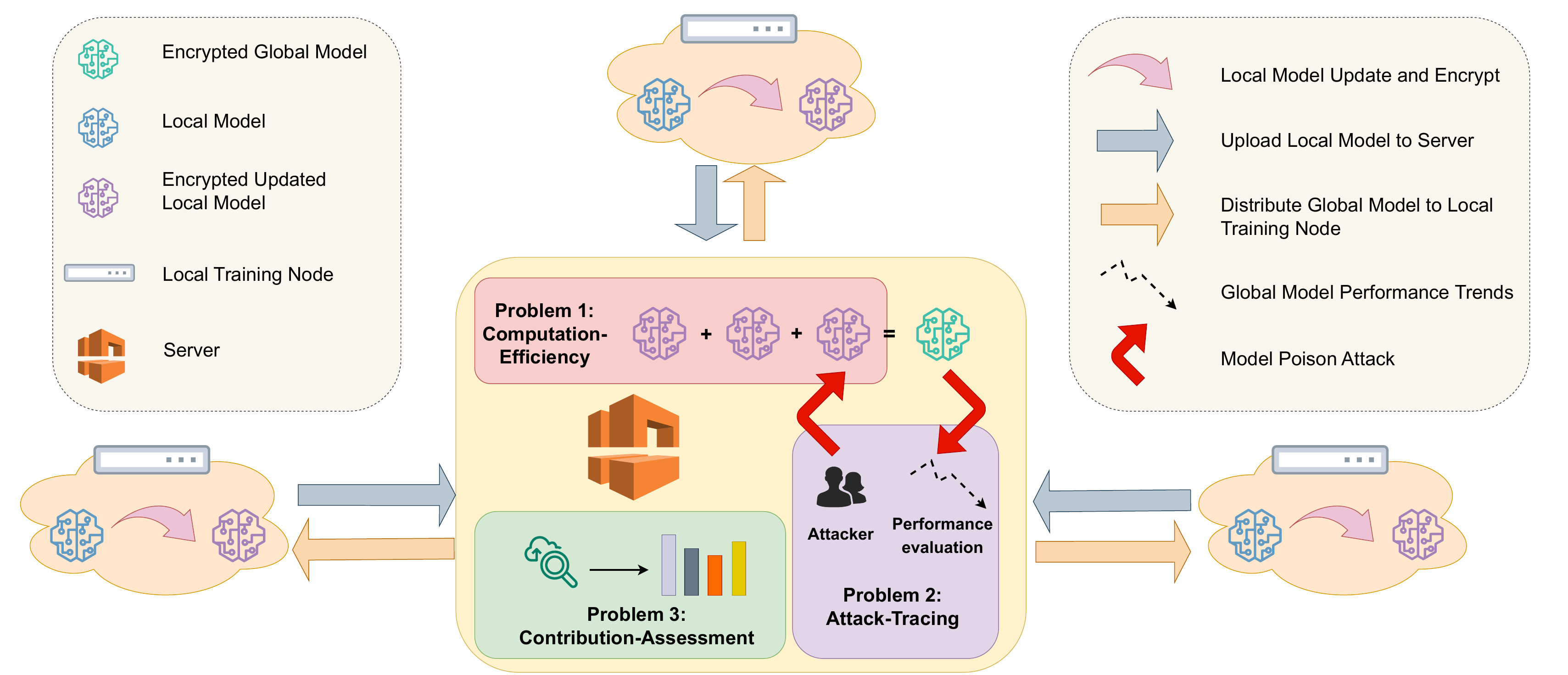}
	\caption{Motivation for this work. Local training nodes update their local models, encrypt them using homomorphic encryption, and send them to the server. The server performs homomorphic computation, which draws three critical problems in global model aggregation: computation-efficiency, attack-tracing, and contribution-assessment. Computation-efficiency implies that homomorphic computation requires substantial computational resources. Attack-tracing refers to the covert nature and difficulty in tracing attacks initiated by local models in an encrypted state. Contribution-assessment refers to the challenge of calculating the contributions made by encrypted local models and providing fair rewards. }
	\label{FLSSM-scenario}
\end{figure*} 
\subsection{Trusted Timestamps}
Trusted timestamps are defined in RFC 3161 \cite{adams2001rfc3161}, essentially binding the hash value of a user's electronic data with an authoritative time source. Based on this, the timestamp server generates a signature, producing an unforgeable timestamp file that proves a specific file or data existed before a certain time. Through hash values and digital signatures, trusted timestamps ensure that files have not been tampered with after the timestamp generation. Trusted timestamps rely on trusted Timestamp Authorities (TSAs), which are typically recognized by governments or industries and have high credibility. Anyone can verify the validity of a timestamp using the TSA's public key, ensuring the transparency and fairness of the process. The main steps are as follows:
\begin{enumerate}
	\item \textbf{Generate Digest}. Calculate a hash of the data for which a timestamp is to be generated, producing a data digest $h$:
	\begin{equation}
		h = H(D)
	\end{equation}
	where $H$ is the hash operation, $D$ is the data.
	\item \textbf{Initiate Timestamp Request}.
	The user sends a request to the Time Stamping Authority (TSA), including the hash value $h$ of the data, to generate a timestamp $\sigma$:
	\begin{equation}
		\sigma = Sign_{TSA}(T)
	\end{equation}
	Hence, $T = (h, time)$ where $h$ is the hash value of the data, and $time$ is the current time, $Sign_{TSA}$ is TSA signature the time and data digest to generate $\sigma$.
	\item \textbf{Return Timestamp}.
	The TSA returns $(T, \sigma)$ to the user.
	\item \textbf{Verify Timestamp}.
	After receiving the timestamp $(T, \sigma)$ returned by the TSA, the user can verify whether the signature $\sigma$ is authentic using the TSA's public key.
	\begin{equation}
		\text{Verify}_{\text{TSA}}(T, \sigma) = \text{True}
	\end{equation}
\end{enumerate}

\section{ Methodology}
\label{Methodology}
In this section, we will provide a detailed introduction of our proposed model. First we would like to present a statement regarding the practical FL privacy security problem. Then, we will provide an overview of the proposed  FL framework. Our method consists of three main parts: Hierarchical Aggregation Mechanism Based on Homomorphic Encryption (HAM),Model Access Control Mechanism Based on Shamir's Secret Sharing (MACM) and Incentive Mechanism Based on Trusted Time Intervals (IMTTI). 
%框架中使用的每个模块将在下面单独的小节中进行描述。
Each module used in the framework will be described in separate subsections below.

%\subsection{ Problem Statement}
%我们的目标是在一个共同的场景中解决一个联邦学习隐私安全问题。该场景涉及两个不同的组织：监管部门和参与公司。两个组织寻求共同完成模型的训练，但监管部门需要对参与公司的本地模型进行监管，防止恶意行为的发生，因此，如何平衡参与公司本地模型的安全性、隐私性和模型效率成为关键问题。Figure \ref{label} depicts 同态加密下联邦学习训练过程。 为了进一步阐述，我们可以将这一场景归纳为3个关键问题：
%\begin{enumerate}
%\item 加密技术效率问题. 现有保护本地模型隐私安全最有效的方法是通过加密技术，对本地模型进行加密后传输和聚合，如同态加密，安全多方计算等，但加密技术需要消耗大量的计算资源，对聚合服务器的计算能力提出了较高要求。
%\item 历史模型访问控制问题. 加密技术如同态加密等可以较好的保护本地模型的隐私，但也为恶意节点发起的模型攻击提供了可趁之机，模型加密后无法被访问和监管，现有安全聚合技术难以在本地模型加密后筛选出恶意模型，攻击者可能发起投毒攻击或后门攻击，破坏联邦学习训练过程。
%\item 激励机制公平性与隐私问题. 联邦学习中需要在不可信环境下由多个公司共同对模型进行训练，因此需要激励机制维护联邦学习模型训练的可持续性。公平的激励机制可以在个体基础上更好的激励公司训练模型，现有研究联邦学习中激励机制贡献公平性的研究可分为基于Shapley值、边际贡献、梯度贡献等,但评估过程中可能需要暴露参与方的部分数据或模型信息，可能造成模型隐私泄露问题。
%\end{enumerate}

\subsection{Problem Statement}
Our goal is to address a federated learning privacy security problem in a common scenario. This scenario involves two different organizations: regulatory authorities and participating clients. Both organizations seek to collaboratively complete model training, but regulatory authorities need to oversee local models of participating clients to prevent malicious behavior. Therefore, how to balance the security, privacy, and efficiency of participating clients' local models becomes a key issue. Figure \ref{FLSSM-scenario} depicts the federated learning training process under homomorphic encryption. To further elaborate, we can summarize this scenario into 3 key problems:

\begin{enumerate}
	\item Encryption technology efficiency problem. The most effective existing method for protecting local model privacy and security is through encryption technologies, such as homomorphic encryption and secure multi-party computation, where local models are encrypted before transmission and aggregation. However, encryption technologies require substantial computational resources, placing high demands on the aggregation server's computational capabilities.
	
	\item Historical model access control problem. Encryption technologies like homomorphic encryption can effectively protect the privacy of local models, but they also provide opportunities for model attacks launched by malicious nodes. Once encrypted, models cannot be accessed or supervised. Existing secure aggregation techniques struggle to filter out malicious models after local model encryption, allowing attackers to potentially launch poisoning attacks or backdoor attacks that disrupt the federated learning training process.
	
	\item Incentive mechanism fairness and privacy problem. Federated learning requires multiple clients to jointly train models in an untrusted environment, necessitating incentive mechanisms to maintain the sustainability of federated learning model training. Fair incentive mechanisms can better motivate clients to train models on an individual basis. Existing research on incentive mechanism contribution fairness in federated learning can be categorized into approaches based on Shapley values, marginal contributions, gradient contributions, etc. However, the assessment process may require exposing some data or model information of the participants, potentially causing model privacy leakage issues.
\end{enumerate}

%\begin{table}[h]
%	\centering
%	\caption{List of Notations}
%	\label{tab:List of Notations}
%	\begin{tabularx}{\columnwidth}{lX}
%		\toprule
%		Notations & Descriptions  \\
%		\midrule
%		FL & Federated Learning \\
%		$h_g$ & 全局模型的hash值 \\
%		$\mathcal{L}_{\kappa}$ & $Ln_{\kappa}$的损失函数 \\	
%		$lm_{\lambda}^{\kappa}$ & $Ln_{\kappa}$在$\lambda$轮次开始时的本地模型参数 \\
%		$lm_{u;\lambda}^{\kappa}$ & $Ln_{\kappa}$在$\lambda$轮次更新后的本地模型参数 \\
%		$lm_{u;\lambda}^{\kappa;\Lambda}$ & $lm_{u;\lambda}^{\kappa}$的第$\Lambda$分片的参数 \\
%		$En_{\iota}^{\lambda,\Lambda}$ & 在第$\lambda$轮次中负责所有本地模型的第$\Lambda$分片模型参数聚合的为$En_{\iota}$\\
%		\bottomrule
%	\end{tabularx}
%\end{table}
\begin{table}[h]
	\centering
	\caption{List of Notations}
	\label{tab:List of Notations}
	\begin{tabularx}{\columnwidth}{lX}
		\toprule
		Notations & Descriptions \\
		\midrule
		FL & Federated Learning \\
		$h_g$ & Hash value of the global model \\
		$\mathcal{L}{\kappa}$ & Loss function of $Ln{\kappa}$ \\
		$lm_{\lambda}^{\kappa}$ & Local model parameters of $Ln_{\kappa}$ at the start of round $\lambda$ \\
		$lm_{u;\lambda}^{\kappa}$ & Updated local model parameters of $Ln_{\kappa}$ after round $\lambda$ \\
		$lm_{u;\lambda}^{\kappa;\Lambda}$ & Parameters of the $\Lambda$-th shard of $lm_{u;\lambda}^{\kappa}$ \\
		$En_{\iota}^{\lambda,\Lambda}$ & $En_{\iota}$ responsible for aggregating the $\Lambda$-th shard of all local model parameters in round $\lambda$ \\
		\bottomrule
	\end{tabularx}
\end{table}

\subsection{ Model Overview}
\begin{figure*}[htbp]
	\centering
	\includegraphics[width=0.9\textwidth]{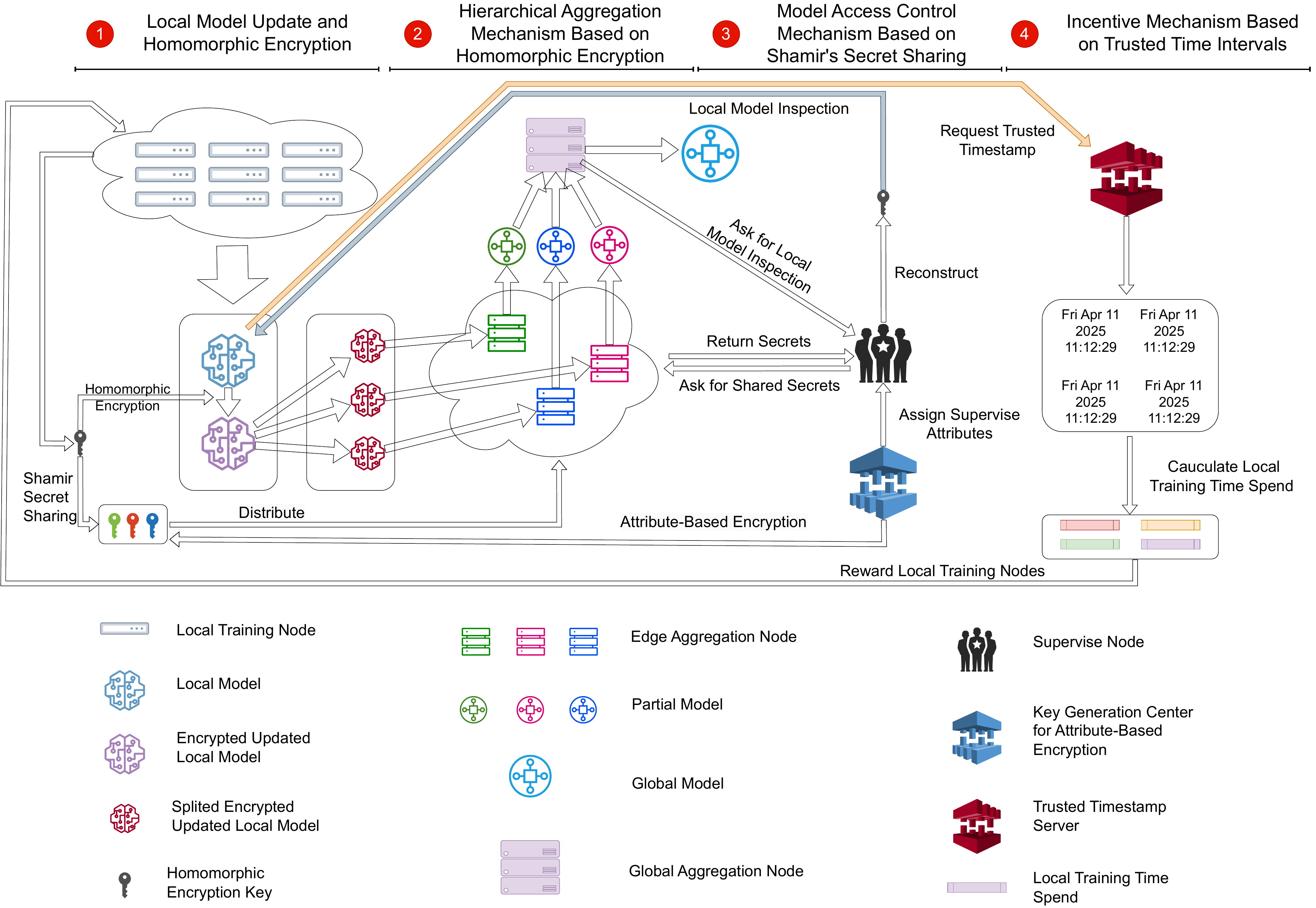}
	\caption{Overview of FLSSM. Federated learning can be broadly divided into two parts: model training and model aggregation. Our research focuses on the second part, where we introduce edge aggregation nodes, supervise nodes, and trusted timestamp servers to improve model aggregation efficiency, enhance security, and provide reliable evidence for incentive mechanisms. }
%	\caption{Overview of FLSSM.联邦学习大致可以分为2个部分：模型训练、模型聚合。我们的研究聚焦于第二部分，通过引入边缘聚合节点,监管节点和可信时间戳服务器加快模型聚合效率，提高安全性，并为激励机制提供可信依据。 }
	\label{FLSSM-structure}
\end{figure*}

%
%As shown in Figure \ref{label}, the proposed model is a  federated learning framework 旨在解决上述3个问题。
%
%该框架主要包括3个模块，下文将对这三个模块进行详细阐述。
%在我们提出的基于同态加密的分布式聚合机制中，本地模型不同层的参数分配至不同的边缘聚合节点进行聚合，从而实现同态加密状态下本地模型的并行聚合，提高全局模型聚合效率，该部分机制is described in Section \ref{sec:homo_aggre}. 历史模型访问控制机制将本地模型的同态加密密钥授权给多个可信节点共同持有，在保护了本地模型存储安全的前提下实现模型监管和恶意行为追溯，这部分内容 is described in Section \ref{sec:shamir}. 
As shown in Figure \ref{FLSSM-structure}, the proposed model is a federated learning framework designed to address the aforementioned three issues.

This framework primarily comprises three modules, which will be elaborated upon in detail below. In our proposed distributed aggregation mechanism based on homomorphic encryption, the parameters of different slices of the local models are distributed to different edge aggregation nodes for aggregation, thereby achieving parallel aggregation of local models under homomorphic encryption and improving the efficiency of global model aggregation. This mechanism is described in Section \ref{sec:homo_aggre}. The historical model access control mechanism authorizes multiple trusted nodes to jointly hold the homomorphic encryption keys of local models. This approach ensures the security of local model storage while enabling model monitoring and malicious behavior traceability. This content is described in Section \ref{sec:shamir}.
%Also, 联邦学习中激励机制的公平性对于联邦学习可持续健康发展至关重要，而现有的基于加密技术的联邦学习隐私保护方法难以对模型贡献进行准确评估。我们设计了一种可信的节点参与积极性评估方法对节点进行奖励，从而在保护本地模型隐私的前提下对参与公司进行公平激励。这部分内容 is described in Section \ref{sec:IMTTI}. 
Furthermore, the fairness of incentive mechanisms in federated learning is crucial for its sustainable and healthy development. However, existing privacy-preserving methods based on encryption technology in federated learning is hard to accurately evaluate model contributions. We have designed a trustworthy node participation activeness evaluation method to reward nodes, thereby providing fair incentives to participating entities while protecting the privacy of local models. This content is described in Section \ref{sec:IMTTI}.
%共识算法旨在保护本地模型隐私的前提下对参与公司进行公平激励。

\subsection{Initialization}
In FLSSM, we denote all nodes' set is $N = {n_1,n_2,\ldots,n_{\theta}},\theta \in \mathbb{N}^+$ is the number of all nodes. $N$ contains global aggregation nodes set, edge aggregation nodes set, local training nodes set. There has three type of nodes in FLSSM:

%\begin{enumerate}
%	\item Global aggregation node($Gn$): The set of global  aggregation nodes is denoted as $GN = \{Gn_1,Gn_2,\ldots,Gn_{\vartheta}\}$, where $\vartheta \in \mathbb{N}^+$.
%	这些节点将边缘聚合节点已经聚合好的各个不同层的本地模型聚合为最终的全局模型.
%	\item Edge aggregation node($En$):
%	The set of edge aggregation node is denoted as $EN = \{En_1,En_2,\ldots,En_{\iota}\}$, where $\iota \in \mathbb{N}^+$.这些节点收到本地模型加密后的某一层的参数后，将同一层的本地模型参数进行初次聚合，然后发送给$Gn$进行最终聚合。
%	\item Local training node($Ln$):The set of local training nodes is denoted as $LN = \{Ln_1,Ln_2,\ldots,Ln_{\kappa}\}$, where $\kappa \in \mathbb{N}^+$.
%	The local training nodes are responsible for training the local models, and encrypt different layer of local model, send encrypted layer to different $En$ for 初次聚合。
%	\item 监管节点 Supervise nodes(Sn): The set of supervise nodes is denoted as $SN = \{Sn_1,Sn_2,\ldots,Sn_{\eta}\}$,where $\eta \in \mathbb{N}^+$. 当模型可能受到攻击需要对本地模型进行校验和审查时，$Gn$会向$Sn$发起监管请求，$Sn$收到监管请求后获取本地模型的加密密钥，对本地模型进行审查。
%\end{enumerate}
\begin{enumerate}
	\item Global aggregation node($Gn$): The set of global  aggregation nodes is denoted as $GN = \{Gn_1,Gn_2,\ldots,Gn_{\vartheta}\}$, where $\vartheta \in \mathbb{N}^+$.
	These nodes aggregate the various slices of local models, which have already been aggregated by the edge aggregation nodes, into the final global model.
	\item Edge aggregation node($En$):
	The set of edge aggregation nodes is denoted as $EN = \{En_1,En_2,\ldots,En_{\iota}\}$, where $\iota \in \mathbb{N}^+$. These nodes receive the encrypted parameters of a specific layer from the local models, perform an initial aggregation of the local model parameters of the same layer, and then send them to $Gn$ for final aggregation.
	\item Local training node($Ln$):The set of local training nodes is denoted as $LN = \{Ln_1,Ln_2,\ldots,Ln_{\kappa}\}$, where $\kappa \in \mathbb{N}^+$.
	The local training nodes are responsible for training the local models, encrypting different slices of the local models, and sending the encrypted slices to different $En$ for initial aggregation.
	\item Supervise nodes($Sn$): The set of supervise nodes is denoted as $SN = \{Sn_1,Sn_2,\ldots,Sn_{\eta}\}$,where $\eta \in \mathbb{N}^+$. When the model may be under attack and local models need to be verified and reviewed, $Gn$ will initiate a inspection request to $Sn$. Upon receiving the inspection request, $Sn$ obtains the encryption keys of the local models to review them.
\end{enumerate}
There are three types of models in FLSSM:
%\begin{itemize}
%	\item Local model ($lm$): The local model set is denoted as $LM = \{lm_{\lambda}^1,lm_{\lambda}^2,\ldots,lm_{\lambda}^{\kappa}\}$, where $\lambda \in \mathbb{N}^+$ represents the global training round, and $\kappa$ represents the node id. $lm_{\lambda}^{\kappa}$ 表示$Ln_{\kappa}$在轮次$\lambda$中更新后的本地模型. 
%	\item Partial model ($pm$): The partial model set is denoted as $PM = \{pm_{\lambda}^1,pm_{\lambda}^2,\ldots,pm_{\lambda}^{\Lambda}\}$, where $\Lambda \in \mathbb{N}^+$ represents the 本地模型平均分配后的第$\Lambda$分片. $En$收到本地训练节点同态加密后的某一层模型参数后，将收到的该层参数通过同态计算进行聚合，得到$pm$.$pm$被发送给$Gn$进行聚合，以密文的形式得到第$\lambda$轮次全局模型。
%	\item Global model ($gm$):The global model set is denoted as $GM = \{gm_{1},gm_{2},\ldots,gm_{\mu}\}$, where $\mu \in \mathbb{N}^+$ represents the global round number. The global model is aggregated by $Gn$ from collected $pm$.
%\end{itemize}
\begin{itemize}
	\item Local model ($lm$): The local model set is denoted as $LM = \{lm_{\lambda}^1,lm_{\lambda}^2,\ldots,lm_{\lambda}^{\kappa}\}$, where $\lambda \in \mathbb{N}^+$ represents the global training round, and $\kappa$ represents the node id. $lm_{\lambda}^{\kappa}$ represents the local model updated by $Ln_{\kappa}$ in round $\lambda$.
	\item Partial model ($pm$): The partial model set is denoted as $PM = \{pm_{\lambda}^1,pm_{\lambda}^2,\ldots,pm_{\lambda}^{\Lambda}\}$, where $\Lambda \in \mathbb{N}^+$ represents the $\Lambda$-th shard after the local model is evenly divided. After $En$ receives the homomorphically encrypted parameters of a certain layer of the local models, it aggregates the received parameters of that layer through homomorphic computation to obtain $pm$. $pm$ is sent to $Gn$ for aggregation to obtain the global model of the $\lambda$-th round in ciphertext.
	\item Global model ($gm$):The global model set is denoted as $GM = \{gm_{1},gm_{2},\ldots,gm_{\mu}\}$, where $\mu \in \mathbb{N}^+$ represents the global round number. The global model is aggregated by $Gn$ from the collected $pm$.
\end{itemize}

%模型整体流程如下：
%\begin{enumerate}
%	\item 聚合节点将全局模型初始化，分发给本地训练节点；同时计算全局模型的hash值，向可信时间戳服务器发送开始计时请求；
%	\item 本地训练节点使用本地数据对模型进行更新，得到更新后的本地模型；
%	\item 本地训练节点将本地模型按照不同的层进行同态加密，发送给对应的边缘聚合节点，同时计算本地模型的hash值，向可信时间戳服务器发送终止计时请求；
%	\item 本地训练节点利用shamir秘密共享技术将同态加密的密钥进行属性基加密后，分为多个部分，分发给监管节点、(可信节点 如代表节点)，设定只有属性为监管节点才能解密该密钥；
%	\item 边缘聚合节点对收到的本地模型层进行聚合，得到$pm$;
%	\item 边缘聚合节点将$pm$发送给$Gn$,$Gn$对接收到的$pm$聚合得到$gm$;
%	\item 激励机制根据可信时间戳服务器收到的计时请求开始和结束的时间计算每个本地训练节点训练所花费的时间，根据训练时间对本地训练节点进行奖励;
%	\item $Gn$将$gm$分发给$Ln$,重复上述步骤直至全局模型收敛。
%\end{enumerate}
The overall model workflow is as follows:
\begin{enumerate}
	\item The aggregation node initializes the global model and distributes it to the local training nodes. Simultaneously, it calculates the hash value of the global model and sends a start timestamp request to the trusted timestamp server.
	\item The local training nodes update the model using their local data, obtaining the updated local models.
	\item The local training nodes homomorphically encrypt the local model according to different slices and send them to the corresponding edge aggregation nodes. Concurrently, they calculate the hash value of the local model and send an end timestamp request to the trusted timestamp server.
	\item The local training nodes utilize Shamir's secret sharing scheme to perform attribute-based encryption on the homomorphic encryption keys, dividing them into multiple parts and distributing them to the supervise nodes (trusted nodes, such as representative nodes), stipulating that only entities with the 'supervise node' attribute can decrypt the keys.
	\item The edge aggregation nodes aggregate the received local model slices to obtain $pm$.
	\item The edge aggregation nodes send $pm$ to $Gn$, and Gn aggregates the received $pm$ to obtain $gm$.
	\item The incentive mechanism calculates the time spent by each local training node on training based on the start and end times received by the trusted timestamp server and rewards the local training nodes accordingly based on their training time.
	\item $Gn$ distributes $gm$ to $Ln$, and the above steps are repeated until the global model converges.
\end{enumerate}

%%伪代码
%\subsection{ 基于同态加密的分布式聚合机制(Hierarchical Aggregation Mechanism Based on Homomorphic Encryption, HAM)}
%\label{sec:homo_aggre}
%FLSSM中，我们提出了一种Hierarchical Aggregation Mechanism Based on Homomorphic Encryption（HAM）使用加密技术保护本地模型的隐私安全。现有使用加密技术保护联邦学习模型隐私的方法包括同态加密、多方安全计算、差分隐私等，其中使用同态加密技术的方法安全性最高，但需要消耗大量计算资源，严重影响联邦学习训练效率。在这项工作中，我们提出了一种分布式聚合机制，
%
%在HAM中我们选择CKKS对模型参数进行加密，因为CKKS支持高效的浮点数运算，适用于机器学习和深度学习等需要高精度计算的场景，尽管存在一定的误差积累问题，但在许多实际应用中可以通过合理的参数设置来控制误差。
%% CKKS算法
%Cheon-Kim-Kim-Song (CKKS)算法\cite{Homomorphic Encryption for Arithmetic of Approximate Numbers}核心思想是通过在复数域上进行同态加密，使得可以在密文上直接执行加法和乘法运算，而不需要解密数据。这种特性使得CKKS算法特别适合于处理数值数据，如浮点数和复数，因此在机器学习、数据分析和隐私保护等领域有广泛的应用。
%
%我们将本地模型进行展开后，根据不同的边缘聚合节点数量平均分配，分别进行同态加密，加密后的模型层分别发送给不同的边缘聚合节点，以模型层为单位并行聚合为$pm$,在保证模型隐私的基础上提高了同态加密的工作效率。这种方法提高了同态加密下模型聚合效率，减少了模型聚合时的同态加密计算所需时间。
% The algorithm details shows in Algorithm \ref{label}.
 \subsection{Hierarchical Aggregation Mechanism Based on Homomorphic Encryption (HAM)}
 \label{sec:homo_aggre}
 In FLSSM, we propose a Hierarchical Aggregation Mechanism Based on Homomorphic Encryption (HAM) to protect the privacy and security of local models using encryption technology. Existing methods for protecting the privacy of federated learning models using encryption technology include homomorphic encryption, secure multi-party computation, and differential privacy. Among these, methods using homomorphic encryption technology offer the highest level of security but require significant computational resources, severely impacting the training efficiency of federated learning. In this work, we propose a distributed aggregation mechanism.
 
 In HAM, we choose CKKS to encrypt the model parameters because CKKS supports efficient floating-point arithmetic, making it suitable for scenarios requiring high-precision calculations such as machine learning and deep learning. Although there is a certain issue of error accumulation, it can be controlled through reasonable parameter settings in many practical applications.
 % CKKS algorithm
 The core idea of the Cheon-Kim-Kim-Song (CKKS) algorithm \cite{cheon2017homomorphic} is to perform homomorphic encryption over complex numbers, allowing direct execution of addition and multiplication operations on ciphertexts without decryption. This characteristic makes the CKKS algorithm particularly suitable for processing numerical data, such as floating-point and complex numbers, and thus it has a wide range of applications in machine learning, data analysis, and privacy protection.
 
 We flatten the local models and then evenly distribute the flattened parameters according to the number of different edge aggregation nodes. Each part is homomorphically encrypted separately, and the encrypted model slices are sent to different edge aggregation nodes. These slices are aggregated in parallel as $pm$ at the edge aggregation nodes. This approach improves the efficiency of homomorphic encryption while ensuring model privacy. This method enhances the model aggregation efficiency under homomorphic encryption and reduces the time required for homomorphic encryption computations during model aggregation.
 The algorithm details are shown in Algorithm \ref{alg:ham}.
 
The main steps are as follows:
\begin{enumerate}
	\item $Ln$ performs local training to obtain the updated local model. Assuming the current $Ln$ id is $\rho$, and the global model training round is $\lambda$, the updated local model of $Ln$ can be denoted as $lm_{u;\lambda}^{\rho}$;
	\begin{equation}
		lm_{u;\lambda}^{\rho} = lm_{\lambda}^{\rho} - \nabla \mathcal{L}_{\rho}(lm_{\lambda}^{\rho})
	\end{equation}
	Here, $\nabla$ is the learning rate, and $\mathcal{L}_{\rho}$ is the loss function of $Ln_{\rho}$:
	\begin{equation}
		\mathcal{L}_{\rho} = \frac{1}{|\mathcal{D}_{\rho}|} \sum_{(\mathbf{x}, y) \in \mathcal{D}_{\rho}} \ell(lm_{\lambda}^{\rho}; \mathbf{x}, y),
	\end{equation}
	Here, $\mathcal{D}_{\rho}$ is the dataset of $Ln_{\rho}$, $(\mathbf{x}, y)$ are the data points in $\mathcal{D}_{\rho}$, and $\ell(lm_{\lambda}^{\rho}; \mathbf{x}, y)$ is the loss value of $lm_{\lambda}^{\rho}$ on the data $\mathbf{x}, y$.
	\item $Ln_{\rho}$ flattens the local model and divides it equally into several shards, then homomorphically encrypts the different model shards. We adopt the CKKS algorithm as the homomorphic encryption algorithm in FLSSM. Assuming the encryption parameters of $Ln_{\rho}$ include the polynomial modulus $q_{\rho}$, polynomial degree $N_{\rho}$, scaling factor $\Delta_{\rho}$, and public-private key pair $(pk_{\rho}, sk_{\rho})$, the process of homomorphically encrypting the $\Lambda$-th shard ($lm_{u;\lambda}^{\rho;\Lambda}$) of $lm_{u;\lambda}^{\rho}$ can be represented as:
	\begin{equation}
		[lm_{u;\lambda}^{\rho;\Lambda}] = \text{CKKS.Encrypt}(pk_{\rho}, \Delta_{\rho} \cdot lm_{u;\lambda}^{\rho;\Lambda})
	\end{equation}
	\item $Ln$ sends the encrypted model parameters of different slices to the corresponding $En$. Assuming $[lm_{u;\lambda}^{\rho;\Lambda}]$ corresponds to $En_{\iota}^{\lambda,\Lambda}$, meaning that in the $\lambda$-th round, $En_{\iota}$ is responsible for aggregating the model parameters of the $\Lambda$-th shard of all local models, we have:
	\begin{equation}
		Ln \rightarrow En_{\iota}^{\lambda,\Lambda} : [lm_{u;\lambda}^{\rho;\Lambda}],
	\end{equation}
	\item $En$ performs homomorphic computation on the received model parameters of the same shard from different $Ln$ to obtain $[pm]$;
	\begin{equation}
		[pm]_{\lambda}^{\Lambda} = \text{CKKS.Add}\left(\frac{1}{\kappa}, \sum_{\rho=1}^{\kappa} [lm_{u;\lambda}^{\rho;\Lambda}]\right)
	\end{equation}
	\item $En$ sends $pm$ to $Gn$, and $Gn$ concatenates the received $[pm]$ to obtain the ciphertext of the global model $[gm]$.
	\begin{equation}
		[gm]_{\lambda} = \text{CKKS.Concat}\left(\{[pm]_{\lambda}^{\Lambda}\}_{\Lambda=1}^{\varrho}\right)
	\end{equation}
	Here, $\varrho$ is the total number of model slices, which also the $En$ number.
	\item $Gn$ distributes $[gm]$ to $Ln$, which decrypts it for the next round of model training.
	\begin{equation}
		\begin{aligned}
			& G_n \xrightarrow{[g_m]} L_i: \text{Train}(\text{CKKS.Decrypt}(sk_i, [g_m])),  \\ & \forall i \in \{1, 2, \ldots, \kappa\}
		\end{aligned}
	\end{equation}
\end{enumerate}
In HAM, we homomorphically encrypt the local model according to different model slices and send them to designated edge aggregation nodes. The edge aggregation nodes aggregate these and send the result to the global aggregation node. Compared to the traditional aggregation mechanism where a single server performs homomorphic encryption and aggregation computation, this reduces the required time and alleviates the computational burden on the server during the aggregation process, thereby improving computational efficiency while ensuring the security of model parameters.

\begin{algorithm}
	\caption{Hierarchical Aggregation Mechanism Based on Homomorphic Encryption (HAM)}
	\label{alg:ham}
	\begin{algorithmic}[1]
	\Require
		\State $Ln_{\rho}$: Local nodes with id $\rho$
		\State $En_{\iota}$: Edge aggregation nodes
		\State $Gn$: Global aggregation node
		\State $\lambda$: Current training round
		\State $\kappa$: Number of local nodes
		\State $\varrho$: Total number of model slices
	\Ensure Global model for next round training
	
	\Function{LocalTraining}{$Ln_{\rho}$, $\lambda$}
		\State Update local model: $lm_{u;\lambda}^{\rho} \gets lm_{\lambda}^{\rho} - \nabla \mathcal{L}_{\rho}(lm_{\lambda}^{\rho})$
		\State Flatten and split $lm_{u;\lambda}^{\rho}$ into $\varrho$ shards
		\For{each shard $\Lambda \in \{1,\ldots,\varrho\}$}
			\State $[lm_{u;\lambda}^{\rho;\Lambda}] \gets \text{CKKS.Encrypt}(pk_{\rho}, \Delta_{\rho} \cdot lm_{u;\lambda}^{\rho;\Lambda})$
			\State Send $[lm_{u;\lambda}^{\rho;\Lambda}]$ to corresponding $En_{\iota}^{\lambda,\Lambda}$
		\EndFor
	\EndFunction
	
	\Function{EdgeAggregation}{$En_{\iota}$, $\lambda$, $\Lambda$}
		\State Collect encrypted shards from all local nodes
		\State $[pm]_{\lambda}^{\Lambda} \gets \text{CKKS.Add}(\frac{1}{\kappa}, \sum_{\rho=1}^{\kappa} [lm_{u;\lambda}^{\rho;\Lambda}])$
		\State Send $[pm]_{\lambda}^{\Lambda}$ to $Gn$
	\EndFunction
	
	\Function{GlobalAggregation}{$Gn$, $\lambda$}
		\State Collect all $[pm]_{\lambda}^{\Lambda}$ from edge nodes
		\State $[gm]_{\lambda} \gets \text{CKKS.Concat}(\{[pm]_{\lambda}^{\Lambda}\}_{\Lambda=1}^{\varrho})$
		\For{each $Ln_i$, $i \in \{1,\ldots,\kappa\}$}
			\State Send $[gm]_{\lambda}$ to $Ln_i$
		\EndFor
	\EndFunction
	
	\Function{LocalDecryption}{$Ln_i$, $[gm]_{\lambda}$}
		\State $gm_{\lambda} \gets \text{CKKS.Decrypt}(sk_i, [gm]_{\lambda})$
		\State Begin next round training with $gm_{\lambda}$
	\EndFunction
	
	\end{algorithmic}
	\end{algorithm}
%%伪代码
%\subsection{ 基于Shamir秘密共享的模型访问控制机制(Model Access Control Mechanism Based on Shamir's Secret Sharing, MACM)}
%\label{sec:shamir}
%在FLSSM中，我们基于Shamir秘密共享算法提出了模型访问控制机制。基于加密机制保护本地模型的隐私安全可以有效的防止数据泄露，但也为恶意节点的攻击提供了可趁之机，服务器无法访问本地模型，只能对加密后的本地模型进行聚合。我们引入了监管节点，用于对本地模型进行检查和校验。对于恶意本地模型，监管节点可以将其踢出当前轮次的全局模型聚合并惩罚其训练节点；当全局模型受到攻击时，监管节点可以访问历史轮次的本地模型，对恶意行为发起者进行追溯。
%
%对本地模型进行同态加密是为了保护本地模型的隐私安全，为了防止监管节点滥用其权力随意访问本地模型造成隐私泄露，我们将$Ln$的同态加密密钥基于Shamir秘密共享技术分为多份，分别存储在本地训练节点、边缘聚合节点、全局聚合节点、监管节点中，当模型受到恶意攻击时，监管节点向保存有密钥分片的节点发起查询请求，申请获得同态加密密钥。Shamir秘密共享技术允许(t,n)访问控制机制，若保存设备掉线，监管节点可以根据其他节点保存的密钥分片进行解密。
%
%同时，同态加密密钥分为多份存储在多个节点中，若恶意节点发起同谋攻击，基于Shamir秘密共享机制的同态加密密钥可能面临被恶意节点恢复的风险。为了防止恶意攻击造成密钥泄露，我们在本地节点将同态加密密钥分片后，额外对分片进行属性基加密，只允许监管节点对分片进行解密，即使恶意节点获得了密钥分片， 也无法将其重组为同态加密密钥，从而保护本地模型的隐私安全性。
%MACM 主要包含两个模块：密钥分发和密钥重组。密钥分发模块中，$Ln$将本地模型的同态加密密钥进行分片和加密后，分发给可信节点；密钥重组模块中，当模型可能受到恶意攻击时，$Sn$向多个可信节点发起密钥重组请求，获得密钥分片后进行解密和重组，得到同态加密的密钥，对本地模型进行审查。
%
%The main steps of 密钥分发 are as follows:
\subsection{Model Access Control Mechanism Based on Shamir's Secret Sharing (MACM)}
\label{sec:shamir}

In FLSSM, we propose a model access control mechanism based on Shamir's secret sharing algorithm. Protecting the privacy and security of local models through encryption effectively prevents data leakage but also creates opportunities for malicious nodes to exploit. Since the server cannot access local models and can only aggregate encrypted local models, we introduce a supervisory node to inspect and validate local models. For malicious local models, the supervisory node can exclude them from the current round of global model aggregation and penalize the corresponding training node. When the global model is under attack, the supervisory node can access historical local models to trace the originator of the malicious behavior.

To protect the privacy of local models, we apply homomorphic encryption. To prevent the supervisory node from arbitrarily accessing local models and causing privacy breaches, we split the homomorphic encryption key of $Ln$ into multiple shares using Shamir's secret sharing technique. These shares are distributed among local training nodes, edge aggregation nodes, global aggregation nodes, and the supervisory node. When the model is subject to a malicious attack, the supervisory node sends queries to nodes holding key shares to request the homomorphic encryption key. Shamir's secret sharing enables a (t,n) access control mechanism, allowing the supervisory node to decrypt the key using shares from other nodes if some devices are offline.

However, storing the homomorphic encryption key across multiple nodes introduces the risk of malicious nodes colluding to reconstruct the key. To mitigate this, after splitting the homomorphic encryption key at local nodes, we further encrypt the shares using attribute-based encryption, ensuring that only the supervisory node can decrypt them. Even if malicious nodes obtain key shares, they cannot reconstruct the homomorphic encryption key, thereby safeguarding the privacy and security of local models.

MACM consists of two main modules: key distribution and key reconstruction. In the key distribution module, $Ln$ splits and encrypts the homomorphic encryption key of the local model and distributes the shares to trusted nodes. In the key reconstruction module, when the model is potentially under a malicious attack, $Sn$ initiates key reconstruction requests to multiple trusted nodes, collects the key shares, decrypts, and reconstructs them to obtain the homomorphic encryption key for reviewing the local model.

The main steps of key distribution are as follows:
%\begin{enumerate}
%	\item $Ln_{\rho}$将同态加密密钥$sk_{\rho}$根据Shamir秘密共享机制分为多片。根据Section \ref{sec:pre-shamir},假设要分片的总数量为$n$, 门限阈值为$t$,随机数为$p$，则构建多项式$f(x)$；
%	\begin{equation}
%		f(sk_{\rho}) = sk_{\rho} + a_1 x + a_2 x^2 + \cdots + a_{t-1} x^{t-1} \pmod{p}
%	\end{equation}
%	选择$n$个不同的数字$x_i,i=\{1,\ldots,n\}$，得到$(x_i,y_i)$作为秘密份额。
%	\item 初始化属性基加密系统，得到主密钥（Master Key）：$Amk$ 和公钥（Public Key）：$Apk$，为每个节点根据属性不同生成特定私钥$Ask$， $Ln_{\rho}$对分片后的密钥进行属性基加密，设置只有监管节点可以解密,假设监管节点的属性集合为$S$,则可得；
%	\begin{equation}
%	CT_i = \mathcal{E}(Apk, (x_i, y_i), S), i \in \{1, \ldots, n \}
%	\end{equation}
%	Hence, $CT_i$ 是$(x_i, y_i)$经过属性基加密后得到的密文。
%	\item $Ln$将加密后的密钥分片$CT_i$分别发送给（代表节点）、（聚合服务器）、监管节点.
%
%\end{enumerate}
%
%当全局模型可能受到投毒攻击或后门攻击导致模型性能下降时，$Gn$会向监管节点发起请求对本地模型进行校验，监管节点可以向持有同态加密密钥分片的节点发起请求申请获取同态加密密钥，对$En$中存储的本地模型进行审查。本地模型的同态加密密钥经过属性基加密，保证了密钥只能被监管节点解密。同时密钥被分为多份，只有满足一定比例的节点同意后监管节点才能获取同态加密密钥，防止了监管节点滥用审查权力随意访问本地模型，造成隐私泄露。监管节点重组同态加密密钥对本地模型审查的过程如下：

\begin{enumerate}
	\item $Ln_{\rho}$ splits the homomorphic encryption key $sk_{\rho}$ into multiple shares using Shamir's secret sharing mechanism. According to Section \ref{sec:pre-shamir}, assuming the total number of shares is $n$, the threshold is $t$, and the random number is $p$, a polynomial $f(x)$ is constructed as follows:
	\begin{equation}
		f(sk_{\rho}) = sk_{\rho} + a_1 x + a_2 x^2 + \cdots + a_{t-1} x^{t-1} \pmod{p}
	\end{equation}
	Select $n$ distinct numbers $x_i, i=\{1,\ldots,n\}$, to obtain the secret shares $(x_i, y_i)$.
	\item Initialize the attribute-based encryption system to generate a master key $Amk$ and a public key $Apk$. Generate a specific private key $Ask$ for each node based on its attributes. $Ln_{\rho}$ encrypts the key shares using attribute-based encryption, configured such that only the supervisory node can decrypt them. Assuming the supervisory node's attribute set is $S$, the encryption is:
	\begin{equation}
		CT_i = \mathcal{E}(Apk, (x_i, y_i), S), i \in \{1, \ldots, n \}
	\end{equation}
	Thus, $CT_i$ is the ciphertext of $(x_i, y_i)$ after attribute-based encryption.
	\item $Ln$ distributes the encrypted key shares $CT_i$ to the representative nodes, aggregation server, and supervisory node.
\end{enumerate}

\begin{algorithm}
	\caption{Model Access Control Mechanism Based on Shamir's Secret Sharing (MACM)}
	\label{alg:macm}
	\begin{algorithmic}[1]
		\Require
		\State $Ln_{\rho}$: Local node with id $\rho$
		\State $Sn$: Supervisory node
		\State $n$: Total number of key shares
		\State $t$: Threshold value
		\State $p$: Large prime number
		\State $S$: Attribute set of supervisory node
		\Ensure Secure model access control
		
		\Function{KeySharing}{$sk_{\rho}$, $n$, $t$, $p$}
		\State Generate random coefficients $a_1,\ldots,a_{t-1}$
		\State Construct polynomial $f(x) = sk_{\rho} + \sum_{i=1}^{t-1} a_i x^i \pmod{p}$
		\For{$i \gets 1$ to $n$}
		\State Select unique $x_i$
		\State Compute $y_i = f(x_i)$
		\State Store share pair $(x_i, y_i)$
		\EndFor
		\Return $\{(x_i, y_i)\}_{i=1}^n$
		\EndFunction
		
		\Function{AttributeBasedEncryption}{shares, $S$}
		\State Generate $(Amk, Apk)$ \Comment{Master key and public key}
		\For{each share pair $(x_i, y_i)$ in shares}
		\State $CT_i \gets \mathcal{E}(Apk, (x_i, y_i), S)$
		\EndFor
		\State Distribute $CT_i$ to trusted nodes
		\Return $\{CT_i\}_{i=1}^n$
		\EndFunction
		
		\Function{KeyReconstruction}{$Sn$}
		\State $shares \gets \emptyset$
		\For{$i \gets 1$ to $t$}
		\State Request $CT_i$ from trusted nodes
		\State $(x_i, y_i) \gets \mathcal{D}(CT_i, SK_M)$ \Comment{Decrypt using ABE}
		\State $shares \gets shares \cup \{(x_i, y_i)\}$
		\EndFor
		\State $sk_{\rho} \gets \sum_{i=1}^t y_i \prod_{j \neq i} \frac{-x_j}{x_i - x_j} \pmod{p}$
		\Return $sk_{\rho}$
		\EndFunction
		
		\Function{ModelInspection}{$sk_{\rho}$}
		\State Request encrypted local model $[m]$ from edge nodes
		\State $m \gets D([m])$ \Comment{Decrypt using reconstructed key}
		\If{IsModelMalicious($m$)}
		\State DeductStake($Ln_{\rho}$)
		\State RemoveFromGlobalModel($m$)
		\If{IsRepeatOffender($Ln_{\rho}$)}
		\State AddToBlacklist($Ln_{\rho}$)
		\EndIf
		\EndIf
		\EndFunction
		
	\end{algorithmic}
\end{algorithm}
When the global model is suspected of being compromised by poisoning or backdoor attacks, leading to performance degradation, $Gn$ requests the supervisory node to validate the local models. The supervisory node can request homomorphic encryption key shares from nodes holding them to review the local models stored in $En$. The homomorphic encryption key shares are protected by attribute-based encryption, ensuring that only the supervisory node can decrypt them. Additionally, the key is split into multiple shares, and the supervisory node can only reconstruct the homomorphic encryption key if a sufficient proportion of nodes consent. This prevents the supervisory node from arbitrarily accessing local models and causing privacy breaches. The process of the supervisory node reconstructing the homomorphic encryption key to review local models is as follows:
%
%\begin{enumerate}
%	\item $Sn$向持有同态加密密钥份额的向代表节点、聚合服务器和监管节点发起密钥重组请求，获取分片后的$CT_i$；
%
%	\item $Sn$获取了满足(t,n)门限的密钥分片，基于属性基加密技术进行解密；
%	\begin{equation}
%		(x_i, y_i) = \mathcal{D}(CT_i, SK_M)
%	\end{equation}
%	\item $Sn$基于Shamir秘密共享技术对密钥分片进行重组，得到同态加密密钥；
%	\begin{equation}
%		sk_{\rho} = \sum_{i=1}^t y_i \prod_{j \neq i} \frac{-x_j}{x_i - x_j} \mod p
%	\end{equation}
%	\item $Sn$向边缘聚合节点发起访问请求，获取本地模型分片，利用同态加密密钥进行解密.
%
%	\begin{equation}
%	m = D([m]) = \frac{L([m]^{\varpi} \pmod{ \sigma^2}}{L(g^{\varpi} \pmod{ \sigma^2})} \pmod{ \sigma}
%	\end{equation}
%
%\end{enumerate}
%
%审查节点发现恶意节点后，会扣除其stake, 并将其本地模型从全局模型中剔除，重新分发全局模型。对于多次发起恶意行为的节点，将被加入黑名单，不再允许参与本地训练。

\begin{enumerate}
	\item $Sn$ sends key reconstruction requests to the representative nodes, aggregation server, and supervisory node holding the homomorphic encryption key shares to obtain the encrypted shares $CT_i$.
	\item $Sn$ collects key shares meeting the (t,n) threshold and decrypts them using attribute-based encryption:
	\begin{equation}
		(x_i, y_i) = \mathcal{D}(CT_i, SK_M)
	\end{equation}
	\item $Sn$ reconstructs the homomorphic encryption key from the key shares using Shamir's secret sharing technique:
	\begin{equation}
		sk_{\rho} = \sum_{i=1}^t y_i \prod_{j \neq i} \frac{-x_j}{x_i - x_j} \mod p
	\end{equation}
	\item $Sn$ sends an access request to the edge aggregation node to obtain the local model shares and decrypts them using the homomorphic encryption key:
	\begin{equation}
		m = D([m]) = \frac{L([m]^{\varpi} \pmod{ \sigma^2}}{L(g^{\varpi} \pmod{ \sigma^2})} \pmod{ \sigma}
	\end{equation}
\end{enumerate}

Upon detecting a malicious node, the supervisory node deducts its stake and removes its local model from the global model aggregation, redistributing the updated global model. Nodes that repeatedly engage in malicious behavior are added to a blacklist and barred from participating in local training.

\subsection{Incentive Mechanism Based on Trusted Time Intervals (IMTTI)}
\label{sec:IMTTI}

Incentive mechanisms are a critical component of federated learning, fostering its sustainable development. Existing studies on federated learning incentives can be categorized into contribution-based, reputation-based, cryptocurrency-based, or blockchain-based approaches. However, these mechanisms primarily target traditional federated learning, evaluating local models in plaintext. In FLSSM, we propose a novel incentive mechanism to reliably assess the participation enthusiasm of local training nodes operating under encrypted conditions, thereby promoting the sustainability of federated learning. Specifically, we leverage a trusted timestamp server to measure the time taken by local training nodes to complete model training. Nodes that complete training faster are considered to have invested more computational resources and are thus rewarded more generously. To prevent nodes from falsely reporting training completion times to gain higher rewards, we mandate that the global aggregation node, responsible for distributing the global model, sends a training start timestamp request to the trusted timestamp server. Local training nodes send an end timestamp request upon completing training. Each timestamp request must include the model's hash value, which is verified against the hash of the aggregated local model during reward allocation to ensure the trustworthiness of the training time.

\begin{algorithm}
	\caption{Incentive Mechanism Based on Trusted Time Intervals (IMTTI)}
	\label{alg:imtti}
	\begin{algorithmic}[1]
		\Require
		\State $Gn$: Global aggregation node
		\State $Ln_{\rho}$: Local node with id $\rho$
		\State $TSA$: Trusted timestamp server
		\State $gm$: Global model
		\State $R_t$: Total reward per round
		\State $\kappa$: Number of local nodes
		\Ensure Reward distribution based on training time
		
		\Function{InitiateTraining}{$Gn$, $gm$}
		\State $h_g \gets H(gm)$ \Comment{Compute hash of global model}
		\State $time \gets \text{CurrentTime}()$
		\State $T \gets (h_g, time)$
		\State $\sigma_g \gets \text{Sign}_{TSA}(T)$ \Comment{Get start timestamp}
		\For{each $Ln_{\rho}$}
		\State Send $(gm, \sigma_g)$ to $Ln_{\rho}$
		\EndFor
		\Return $\sigma_g$
		\EndFunction
		
		\Function{LocalTraining}{$Ln_{\rho}$, $gm$}
		\State Train local model $lm_{\lambda}^{\rho}$ using $gm$
		\State Shard and encrypt local model
		\State $h_l^{\rho} \gets H(lm_{\lambda}^{\rho})$ \Comment{Compute hash of local model}
		\State Send $h_l^{\rho}$ to $TSA$
		\State $\sigma_l^{\rho} \gets \text{Sign}_{TSA}(h_l^{\rho})$ \Comment{Get end timestamp}
		\Return $(lm_{\lambda}^{\rho}, h_l^{\rho}, \sigma_l^{\rho})$
		\EndFunction
		
		\Function{CalculateTimeInterval}{$TSA$, $\sigma_g$, $\sigma_l^{\rho}$}
		\State $\sigma_d^{\rho} \gets \sigma_l^{\rho} - \sigma_g$ \Comment{Calculate time interval}
		\Return $\sigma_d^{\rho}$
		\EndFunction
		
		\Function{CalculateRewards}{$\{\sigma_d^{\rho}\}_{\rho=1}^{\kappa}$, $R_t$}
		\For{each $Ln_{\rho}$}
		\State $\mathfrak{C}_{\rho} \gets e^{-0.1\sigma_d^{\rho}}$ \Comment{Calculate contribution}
		\EndFor
		\State $total\_contribution \gets \sum_{\rho=1}^{\kappa} \mathfrak{C}_{\rho}$
		\For{each $Ln_{\rho}$}
		\State $R^{\rho} \gets \frac{\mathfrak{C}_{\rho}}{total\_contribution} R_t$ \Comment{Calculate reward}
		\EndFor
		\Return $\{R^{\rho}\}_{\rho=1}^{\kappa}$
		\EndFunction
		
		\Function{VerifyAndDistributeRewards}{$Gn$}
		\For{each $Ln_{\rho}$}
		\State Verify $h_l^{\rho}$ matches received encrypted model
		\If{verification successful}
		\State Distribute reward $R^{\rho}$ to $Ln_{\rho}$
		\Else
		\State Skip reward for $Ln_{\rho}$
		\EndIf
		\EndFor
		\EndFunction
		
	\end{algorithmic}
\end{algorithm}

The main steps of IMTTI are as follows:

\begin{enumerate}
	\item The global aggregation node distributes the global model to $Ln$, computes the hash value $h_g$ of the global model, and sends $h_g$ to the trusted timestamp server to mark the training start time:
	\begin{equation}
		h_g = H(gm)
	\end{equation}
	\begin{equation}
		T = (h_g,time)
	\end{equation}
	\begin{equation}
		\sigma_g = Sign_{TSA}(T)
	\end{equation}
	Hence, $Sign_{TSA}$ denotes the trusted timestamp server, $\sigma_g$ represents the start time of local model training, and $time$ is the current time.
	\item Upon receiving $gm$, the local model $Ln_{\rho}$ conducts training. After training is complete, it follows step 2 in Section \ref{sec:homo_aggre} to shard and encrypt the local model. It then computes the hash values of the encrypted local model and sends these hash values to the trusted timestamp server to initiate an end timestamp request:
	\begin{equation}
		h_l^{\rho} = H(lm_{\lambda}^{\rho})
	\end{equation}
	\item Upon receiving the hash values of the local model from $Ln_{\rho}$, the trusted timestamp server generates a trusted timestamp $h_l^{\rho}$ and calculates the time difference between the end and start times, producing a trusted time interval $\sigma_d^{\rho}$:
	\begin{equation}
		\sigma_l^{\rho} = Sign_{TSA}(h_l^{\rho})
	\end{equation}
	\begin{equation}
		\sigma_d^{\rho} = \sigma_l^{\rho} - \sigma_g
	\end{equation}
	\item The incentive mechanism rewards local models based on their trusted time intervals. For $Ln_{\rho}$, the total reward per round $R_t$ is fixed, and the contribution of $Ln_{\rho}$ is calculated as:
	\begin{equation}
		\mathfrak{C}_{\rho} = e^{-0.1 \sigma_d^{\rho}}
	\end{equation}
	The reward received by $Ln_{\rho}$ is then expressed as:
	\begin{equation}
		R^{\rho} = \frac{\mathfrak{C}_{\rho}  } {\sum_{\rho = 1}^{\kappa} \mathfrak{C}_{\rho}} R_t
		\label{Rrho}
	\end{equation}
\end{enumerate}

%\section{ Security Analysis}
\section{ Performance Evaluation}
\label{Performance-Evaluation}
In this section, we conduct experiments and ablation studies on two public datasets to validate our model.
% 我们首先在\ref{Environment}和\ref{Datasets}介绍了实验环境和数据集。Next, 在\ref{Comparative-Experiment}和\ref{Ablation-Experiment}中报告了有关三个模块的实验结果。
We first introduce the experimental environment and datasets in Sections \ref{Environment} and \ref{Datasets}. Next, we report the experimental results for the three modules in Sections \ref{Comparative-Experiment} and \ref{Ablation-Experiment}.

\subsection{ Environment}
\label{Environment}

Our experiments were conducted on a CentOS server, using PyTorch version 2.3.0, CUDA version 12.4, with a system equipped with 396GB of RAM and a 1.8TB hard disk, the GPUs in the server are Quadro RTX 5000 (with a VRAM of 16GB). We use dirichlet distribution to split dataset, the alpha is set to 0.5.

\begin{table}[h]
	\centering
	\caption{Simulation Prameters }
	\label{tab:Simulation Prameters}
	\begin{tabularx}{\columnwidth}{lc}
		\toprule
		Parameters & Value \\
		\midrule
		Learning rate & 0.001 \\
		Batchsize & 64 \\
		Alpha of data Non-IID distribution & 0.5 \\
		Number of Communication Round & 100 \\
		Number of Supervise Node & 1 \\
		Epoch in each round & 5 \\
		Total reward in one round ($R_t$) & 10 \\
		Threshold of Shamir Secret Shares & 3 \\
		Number of Shamir Secret Shares & 5 \\
		Malicious attack & Sign\_Flipping Attack\\
		\bottomrule
	\end{tabularx}
\end{table}
\subsection{ Datasets}
\label{Datasets}
%3个数据集：FEMNIST，CIFAR10，SVHN
\begin{itemize}
%	\item Fashion-MNIST: Fashion-MNIST数据集由70,000张各种时尚产品的正面图像组成,可分为10类, 所有图像均为灰度图像，尺寸为28 × 28像素。训练集包含60,000张图像，测试集包含10,000张图像。
\item CIFAR-10 \cite{krizhevsky2009learning}: The CIFAR-10 dataset contains 3-channel RGB color images across 10 categories, with each image sized at 32 × 32 pixels. Each category includes 6,000 images, resulting in a total of 50,000 training images and 10,000 test images.

\item Fashion-MNIST \cite{xiao2017fashion}: The Fashion-MNIST dataset consists of 70,000 frontal images of various fashion products, categorized into 10 classes. All images are grayscale with dimensions of 28 × 28 pixels. The training set contains 60,000 images, while the test set contains 10,000 images.

%CIFAR10: CIFAR10 共包含10 个类别的3 通道 RGB 彩色图片,每个图片的尺寸为32 × 32 ，每个类别有6000个图像，数据集中一共有50000 张训练图片和10000 张测试图片。
\end{itemize}
%\subsection{Baselines}
%\label{Baselines}
%5个baseline
\subsection{ Comparative Experiment}
\label{Comparative-Experiment}
\subsubsection{Computation-efficiency Experiment}
\label{computation-efficiency}

\begin{figure*}[htbp]
	\centering
	\begin{subfigure}[b]{0.31\textwidth}
		\centering
  \captionsetup{justification=centering}
		\includegraphics[width=\textwidth]{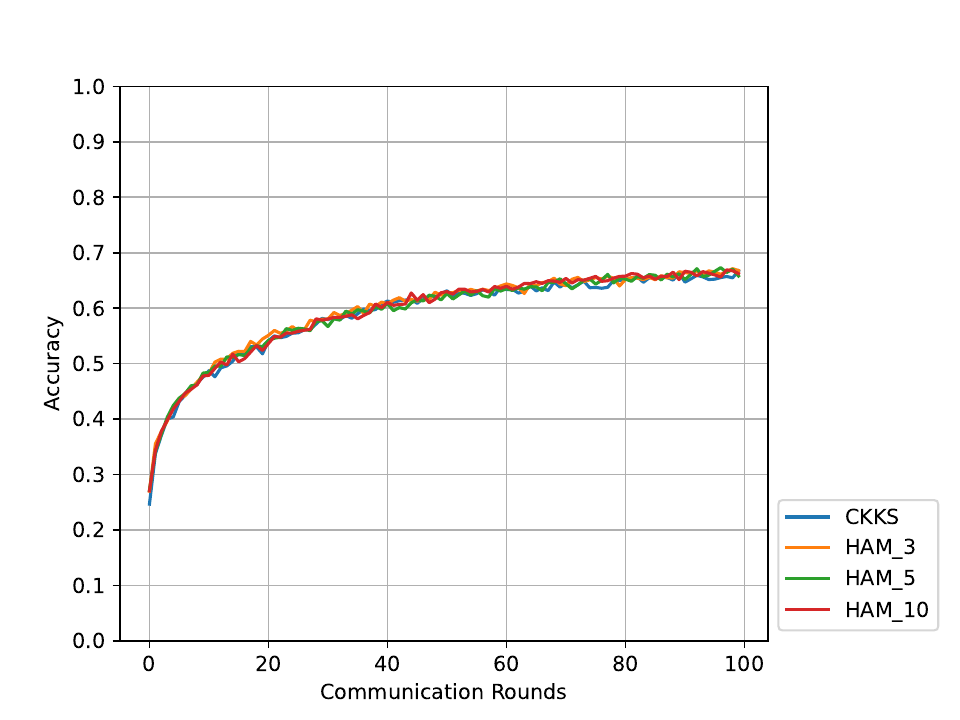}
		\caption{Global Model Accuracy, \\CIFAR10, $Ln$=10}
		\label{GACCIFAR10Ln10}
	\end{subfigure} 
	\begin{subfigure}[b]{0.31\textwidth}
		\centering
  \captionsetup{justification=centering}
		\includegraphics[width=\textwidth]{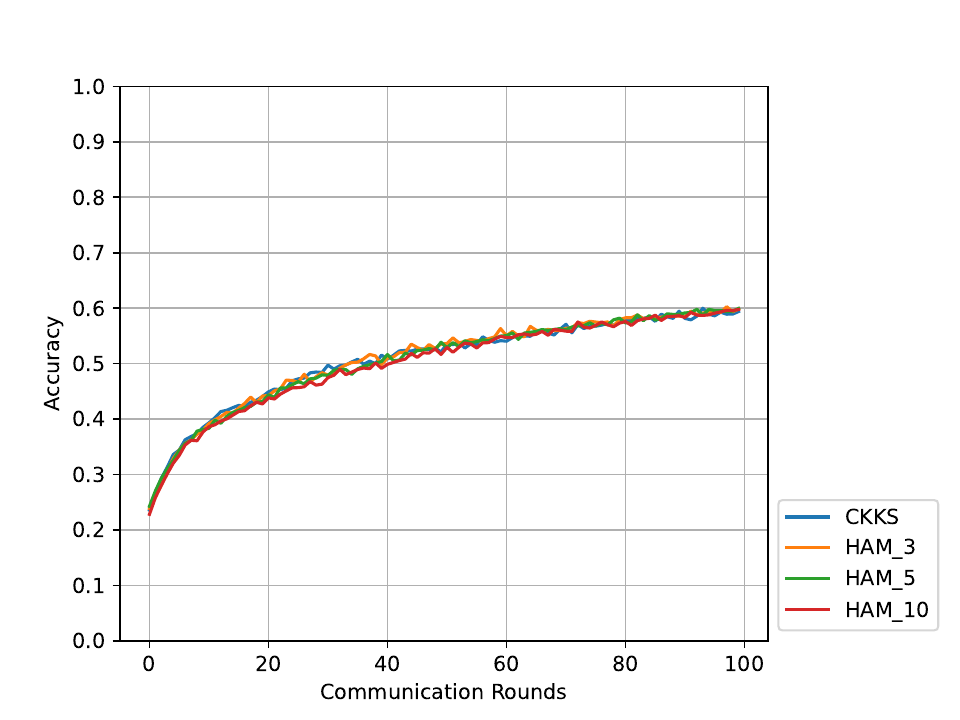}
		\caption{Global Model Accuracy, \\CIFAR10, $Ln$=20}
		\label{GACCIFAR10Ln20}
	\end{subfigure} 
	\begin{subfigure}[b]{0.31\textwidth}
	\centering
  \captionsetup{justification=centering}
	\includegraphics[width=\textwidth]{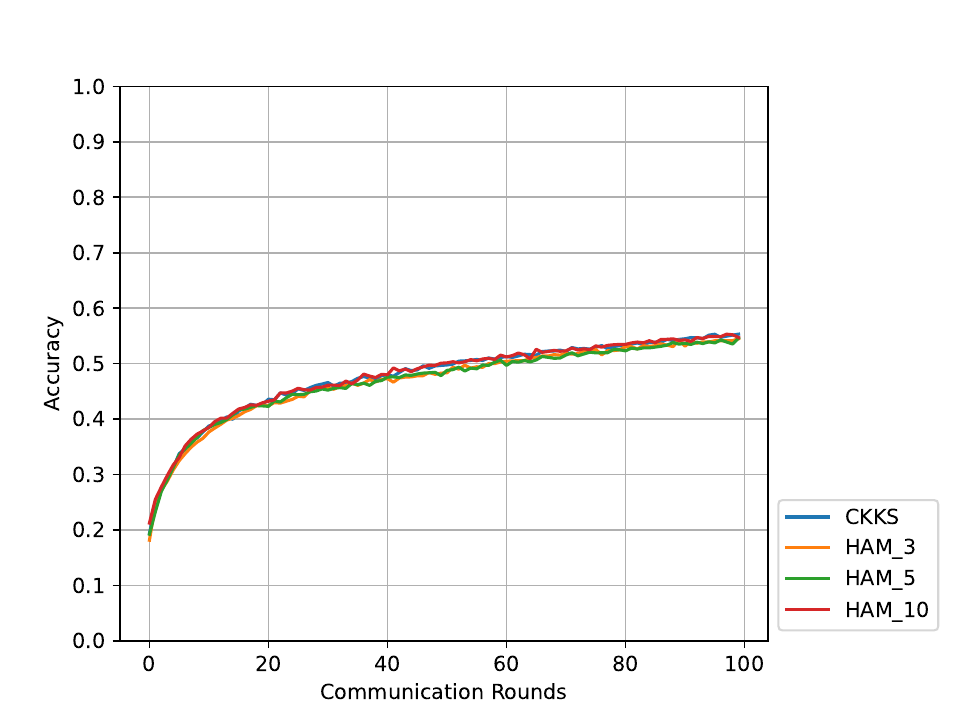}
	\caption{Global Model Accuracy, \\CIFAR10, $Ln$=50}
	\label{GACCIFAR10Ln50}
	\end{subfigure}

	\begin{subfigure}[b]{0.31\textwidth}
		\centering
  \captionsetup{justification=centering}
		\includegraphics[width=\textwidth]{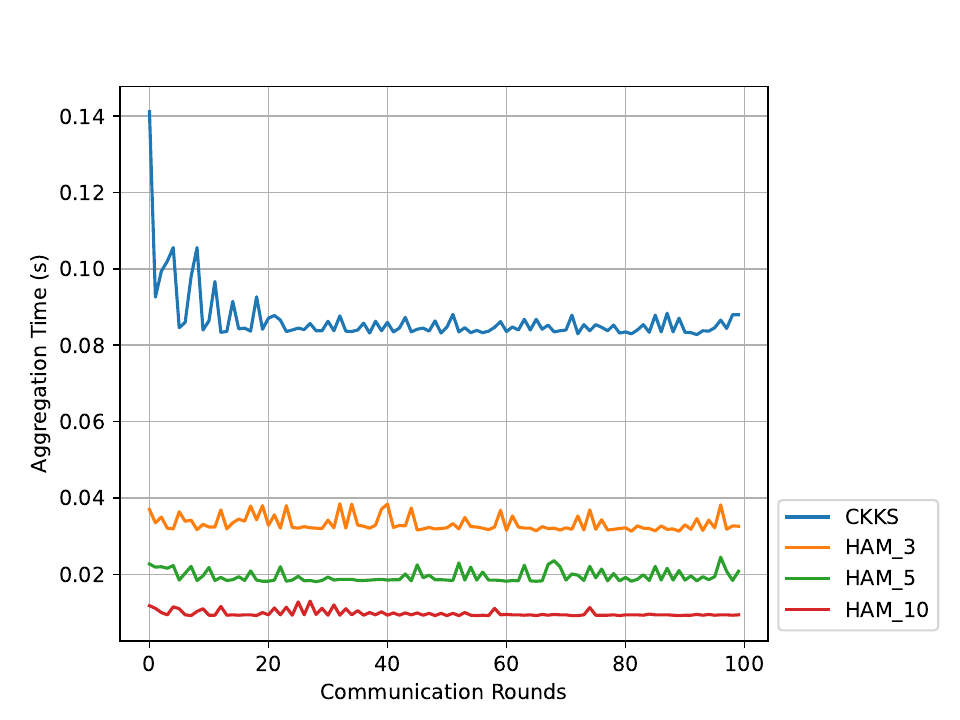}
		\caption{Aggregation Time, \\CIFAR10, $Ln$=10}
		\label{ATCIFAR10Ln10}
	\end{subfigure} 
	\begin{subfigure}[b]{0.31\textwidth}
		\centering
  \captionsetup{justification=centering}
		\includegraphics[width=\textwidth]{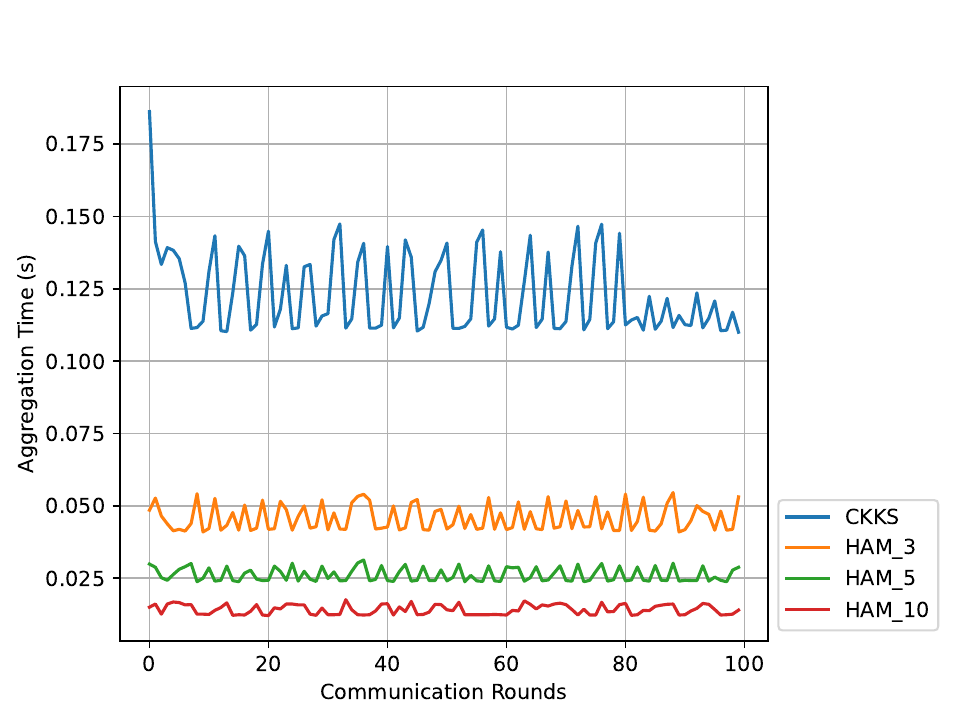}
		\caption{Aggregation Time, \\CIFAR10, $Ln$=20}
		\label{ATCIFAR10Ln20}
	\end{subfigure} 
	\begin{subfigure}[b]{0.31\textwidth}
		\centering
  \captionsetup{justification=centering}
		\includegraphics[width=\textwidth]{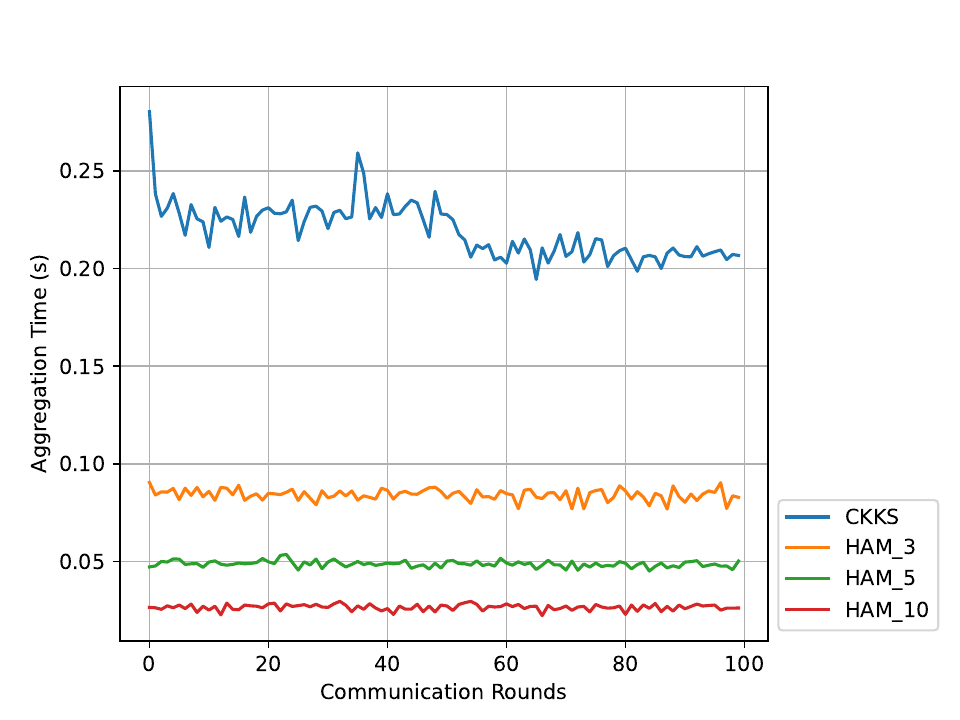}
		\caption{Aggregation Time, \\CIFAR10, $Ln$=50}
		\label{ATCIFAR10Ln50}
	\end{subfigure}

	\begin{subfigure}[b]{0.31\textwidth}
		\centering
  \captionsetup{justification=centering}
		\includegraphics[width=\textwidth]{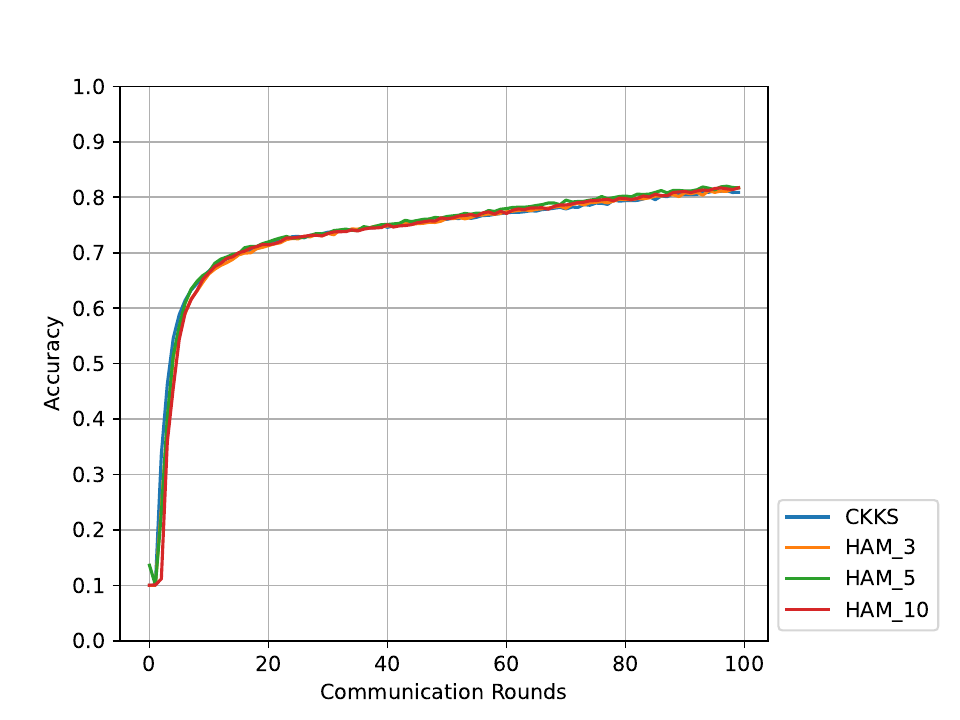}
		\caption{Global Model Accuracy, \\ Fashion-MNIST, $Ln$=10}
		\label{GACFASHIONMNISTLn10}
	\end{subfigure} 
	\begin{subfigure}[b]{0.31\textwidth}
		\centering
  \captionsetup{justification=centering}
		\includegraphics[width=\textwidth]{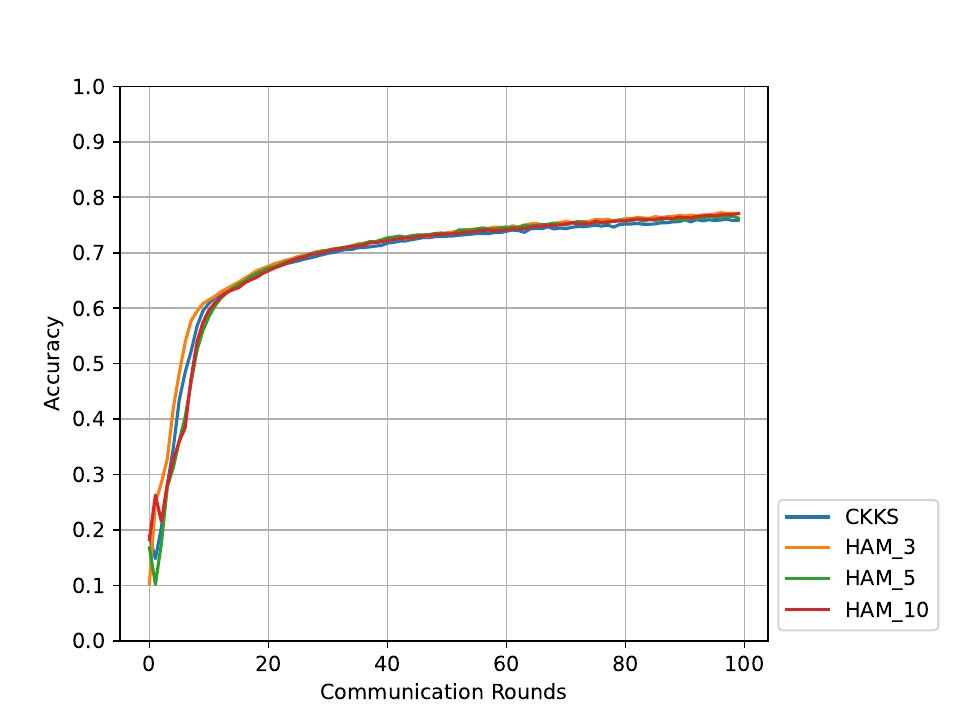}
		\caption{Global Model Accuracy, \\ Fashion-MNIST, $Ln$=20}
		\label{GACFASHIONMNISTLn20}
	\end{subfigure} 
	\begin{subfigure}[b]{0.31\textwidth}
		\centering
  \captionsetup{justification=centering}
		\includegraphics[width=\textwidth]{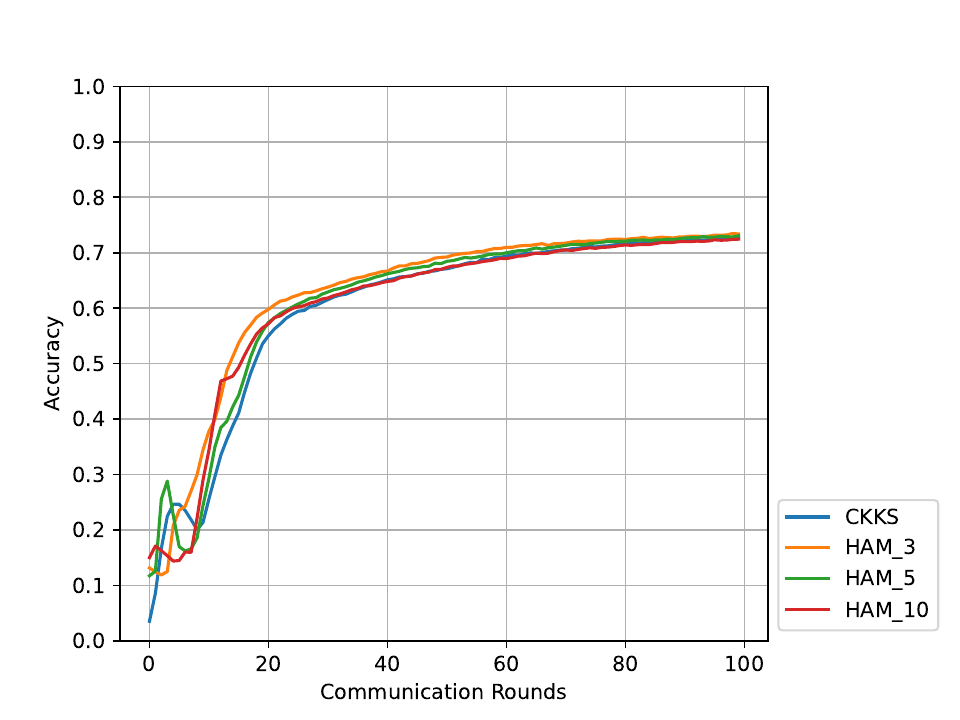}
		\caption{Global Model Accuracy, \\ Fashion-MNIST, $Ln$=50}
		\label{GACFASHIONMNISTLn50}
	\end{subfigure}

	\begin{subfigure}[b]{0.31\textwidth}
		\centering
  \captionsetup{justification=centering}
		\includegraphics[width=\textwidth]{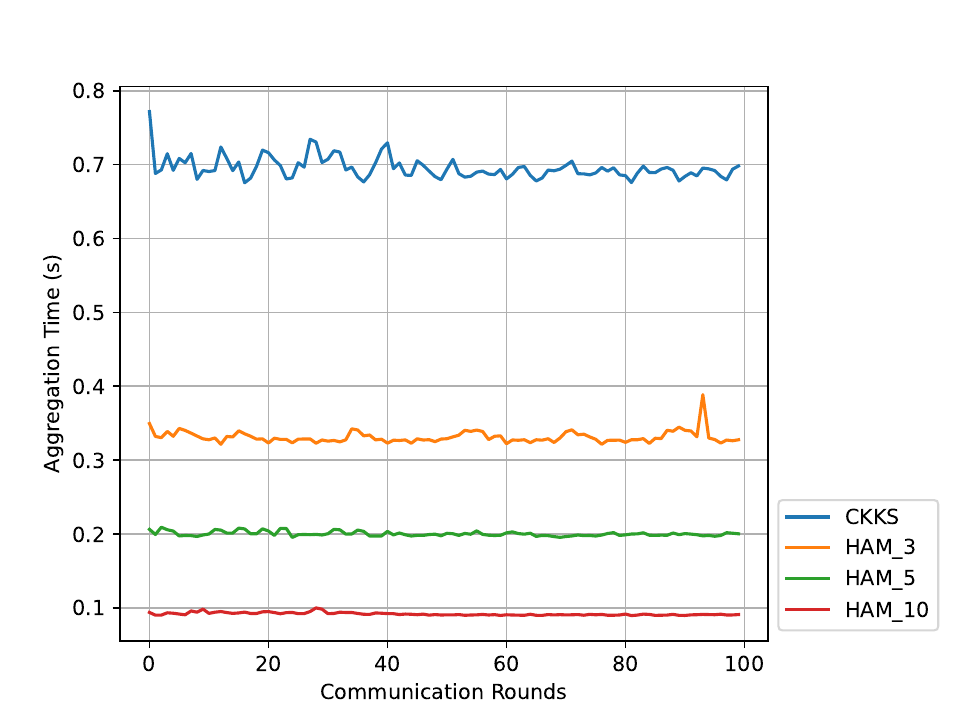}
		\caption{Aggregation Time, \\ Fashion-MNIST, $Ln$=10}
		\label{ATFASHIONMNISTLn10}
	\end{subfigure} 
	\begin{subfigure}[b]{0.31\textwidth}
		\centering
  \captionsetup{justification=centering}
		\includegraphics[width=\textwidth]{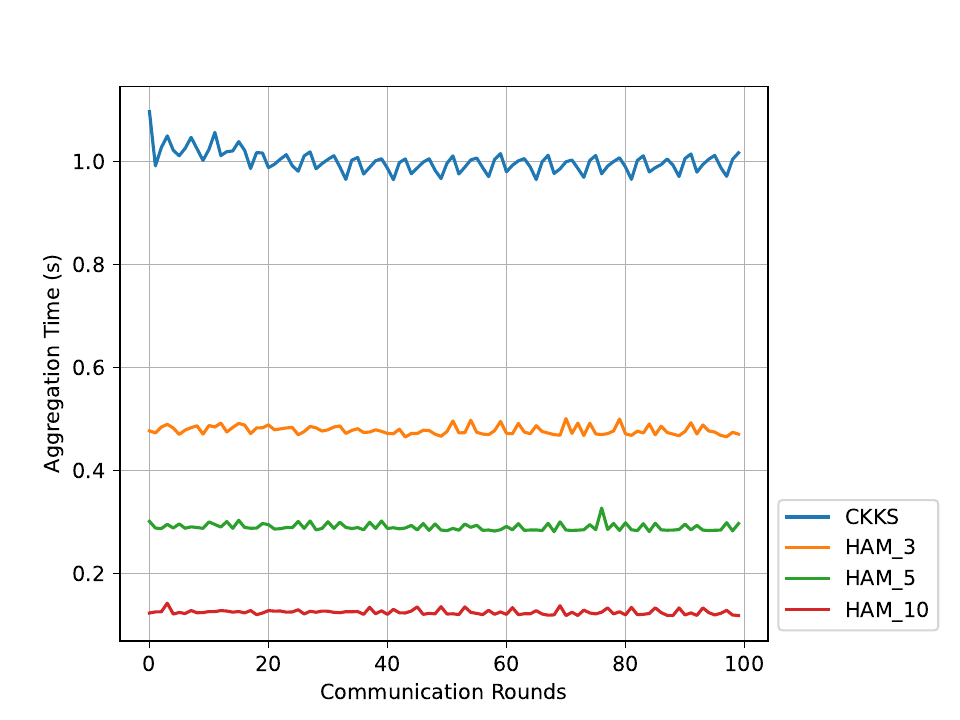}
		\caption{Aggregation Time, \\ Fashion-MNIST, $Ln$=20}
		\label{ATFASHIONMNISTLn20}
	\end{subfigure} 
	\begin{subfigure}[b]{0.31\textwidth}
		\centering
  \captionsetup{justification=centering}
		\includegraphics[width=\textwidth]{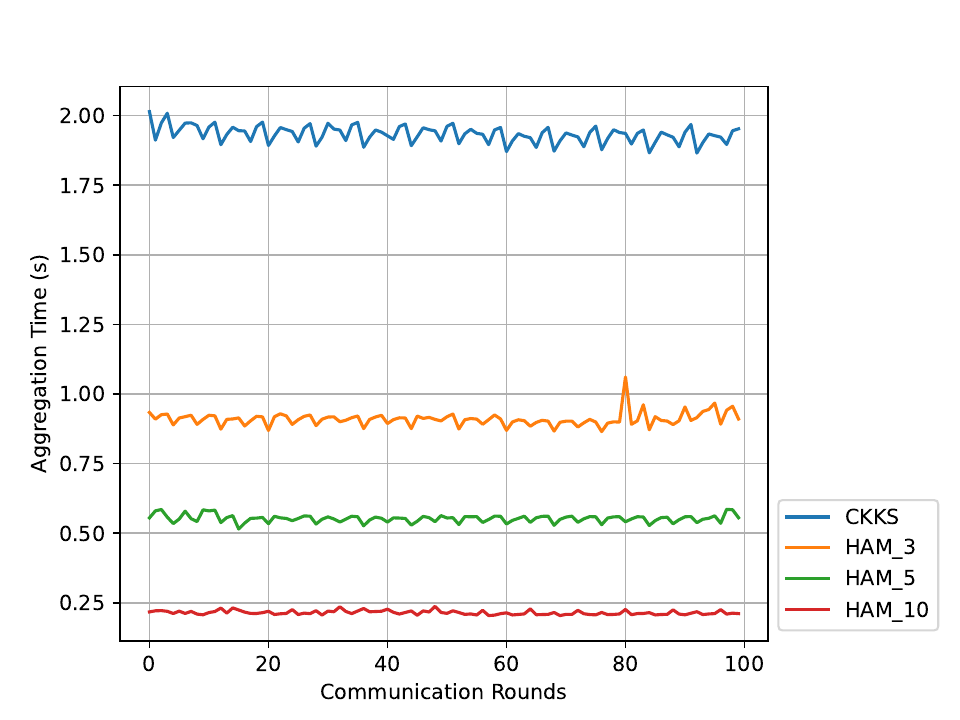}
		\caption{Aggregation Time, \\ Fashion-MNIST, $Ln$=50}
		\label{ATFASHIONMNISTLn50}
	\end{subfigure}

	\caption{
	Accuracy of HAM and CKKS algorithms for CIFAR10 and Fashion-MNIST classification tasks. Figures \ref{GACCIFAR10Ln10}-\ref{GACCIFAR10Ln50} present the global model accuracy on CIFAR10 dataset for local training node ($Ln$) counts  in \{10, 20, 50\} and Edge Aggregation Nodes ($En$) counts in \{3, 5, 10\}. Figures \ref{ATCIFAR10Ln10}-\ref{ATCIFAR10Ln50} show the global model aggregation time on CIFAR10 dataset with the same parameter settings.
	Similarly, Figures \ref{GACFASHIONMNISTLn10}-\ref{GACFASHIONMNISTLn50} illustrate global model accuracy on the Fashion-MNIST dataset, and Figures \ref{ATFASHIONMNISTLn10}-\ref{ATFASHIONMNISTLn50} present the global model aggregation time on Fashion-MNIST dataset under the same parameter configurations.
}
%	\caption{Accuracy of HAM, CKKS for CIFAR10 and Fashion-MNIST task. \ref{GACCIFAR10Ln10}-\ref{GACCIFAR10Ln50}: global model accuracy in CIFAR10 dataset for $Ln$ in \{10, 20, 50\} and $En$ in \{3, 5, 10\}. \ref{ATCIFAR10Ln10}-\ref{ATCIFAR10Ln50}: global model aggregation time in CIFAR10 dataset for $Ln$ in \{10, 20, 50\} and $En$ in \{3, 5, 10\}.
%	\ref{GACFASHIONMNISTLn10}-\ref{GACFASHIONMNISTLn50} is global model accuracy in Fashion-MNIST dataset, and \ref{ATFASHIONMNISTLn10}-\ref{ATFASHIONMNISTLn50} is global model aggregation time in Fashion-MNIST dataset. }
	\label{GACAccuracy}
\end{figure*}

%
%图\ref{GACAccuracy}在CIFAR10 和Fashion-MNIST 两个公开数据集下测试了我们提出的HAM和不使用边缘聚合节点而是直接聚合的情况下聚合所需时间之间的关系，以及模型准确率的对比。其中，``CKKS"表示采用CKKS同态加密算法，不设置边缘聚合节点而是直接由全局聚合节点（Gn）进行聚合。``HAM"表示采用HAM聚合算法，``HAM-3"表示3个边缘聚合节点，``HAM-5"表示5个边缘聚合节点，``HAM-10"表示10个边缘聚合节点。较低的模型聚合时间表示较高的计算效率，当训练节点数量相同时，边缘聚合节点越多，聚合时间越少，聚合效率越高。此外，它表明对于两个数据集，我们所提出的方案都可以通过使用边缘聚合节点有效提高加密模型的同态计算效率。这揭示了边缘聚合节点可以有效提高加密模型的聚合效率，归因于由于加密后的本地模型参数被平均分配给不同的边缘聚合节点，因此边缘聚合节点数量越多，每个节点需要聚合的参数越少。当训练节点数量不同时，聚合时间分布相似。对于相同训练节点数量的情况下，模型准确率相似，不同训练节点数量的情况下，训练节点数量越多，模型准确率会减小。这表明我们提出的方案在提高加密模型计算效率的同时，并未对模型性能产生较大影响。同时，边缘聚合节点使得模型参数分开存储而不是存储在一个中心服务器中，增强了模型参数的存储安全性。
Figure \ref{GACAccuracy} evaluates the relationship between aggregation time and model accuracy under two public datasets, CIFAR10 and Fashion-MNIST, comparing our proposed HAM scheme with a baseline that aggregates directly without edge aggregation nodes. Here, ``CKKS" refers to the CKKS homomorphic encryption algorithm, where aggregation is performed directly by the global aggregation node (Gn) without edge aggregation nodes. ``HAM" denotes the HAM aggregation algorithm, with ``HAM-3" indicating 3 edge aggregation nodes, ``HAM-5" indicating 5 edge aggregation nodes, and ``HAM-10" indicating 10 edge aggregation nodes. Lower model aggregation time indicates higher computational efficiency. When the number of training nodes is the same, a greater number of edge aggregation nodes results in shorter aggregation times and higher aggregation efficiency. Furthermore, the results demonstrate that for both datasets, our proposed scheme effectively enhances the homomorphic computation efficiency of encrypted models by utilizing edge aggregation nodes. This improvement is attributed to the fact that encrypted local model parameters are evenly distributed across different edge aggregation nodes; thus, more edge aggregation nodes mean fewer parameters for each node to aggregate. When the number of training nodes varies, the aggregation time distribution remains similar. For the same number of training nodes, model accuracy is comparable. However, when the number of training nodes increases, model accuracy slightly decreases. This indicates that our proposed scheme improves the computational efficiency of encrypted models without significantly impacting model performance. Additionally, edge aggregation nodes enable model parameters to be stored separately rather than on a single central server, enhancing the storage security of model parameters.

\subsection{Ablation Experiment}
\label{Ablation-Experiment}

\subsubsection{ Attack-tracing Experiment}
\begin{figure*}[htbp]
	\centering
	\begin{subfigure}[b]{0.48\textwidth}
		\centering
		\captionsetup{justification=centering}
		\includegraphics[width=\textwidth]{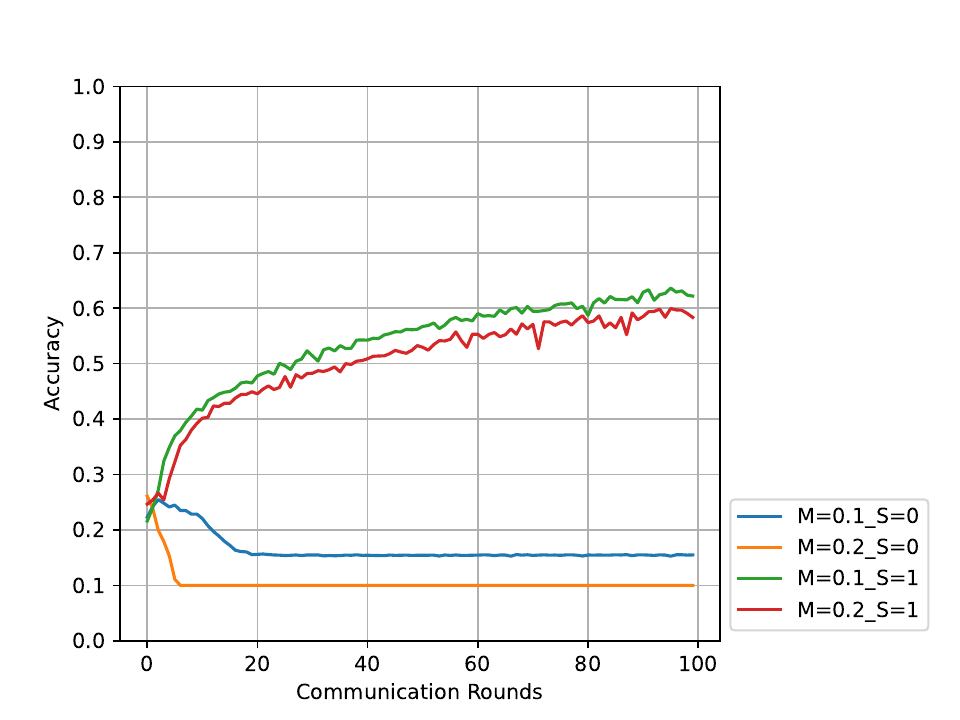}
		\caption{ CIFAR10, $Ln$=10}
		\label{GACMCIFAR10Ln10}
	\end{subfigure} 
	\begin{subfigure}[b]{0.48\textwidth}
		\centering
		\captionsetup{justification=centering}
		\includegraphics[width=\textwidth]{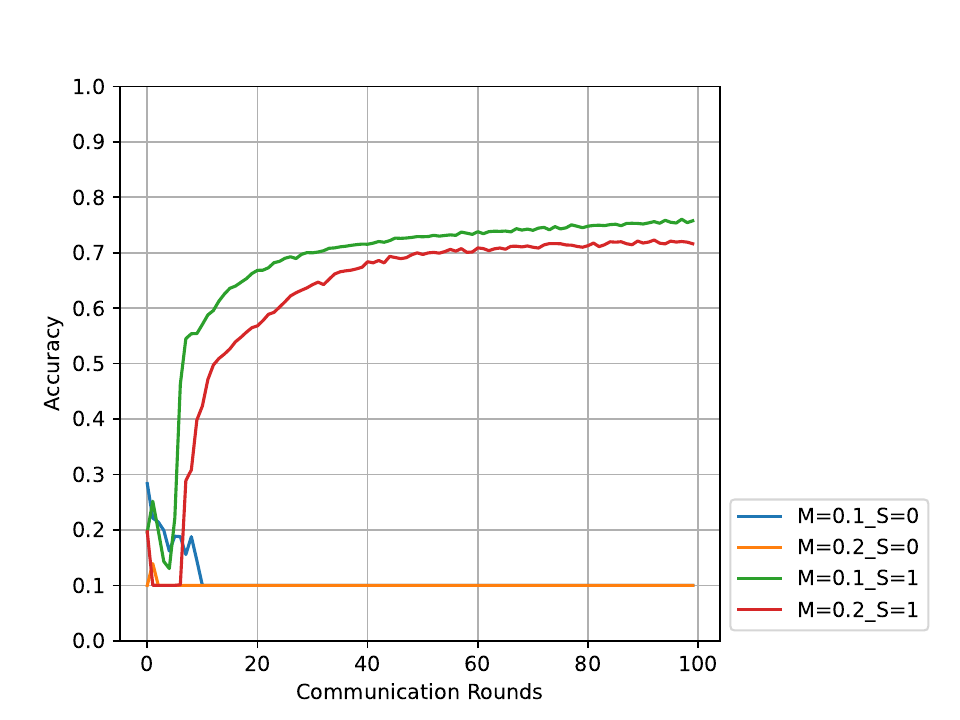}
		\caption{ Fashion-MNIST, $Ln$=10}
		\label{GACMFASHIONMNISTLn10}
	\end{subfigure}

	\caption{
	Global Model Accuracy  of HAM algorithm under Malicious Attack for CIFAR10 and Fashion-MNIST classification tasks. ``M" in the legend represents the malicious nodes ratio in local training nodes, and ``S" represents whether the Supervise Node setting to inspection or not. ``S=1": inspection; ``S=0": not inspection.
	}
	\label{GACMAccuracy}
\end{figure*}

%图\ref{GACMAccuracy}展示了本地训练节点（Ln）中存在不同比例的恶意节点的情况下，我们提出的MACM机制中的监管节点对本地模型的审查与模型表现之间的关系, with number of $Ln$ is 10。\ref{GACMCIFAR10Ln10}展示了在CIFAR10 数据集中，恶意节点比例分别为\{0.1, 0.2\}时，监管节点审查本地模型对模型性能的影响。当监管节点不审查本地模型时（S=0），模型性能很快下降到了0.1-0.2之间，并且恶意模型比例越高模型性能越低。当监管节点审查本地模型时（S=1），模型性能相比于不审查本地模型表现出了较大提升。
%\ref{GACMFASHIONMNISTLn10}展示了在Fashion-MNIST数据集中监管节点对模型性能的影响。与图\ref{GACMAccuracy}相同的是监管节点对本地模型的审查会有效提高模型表现。
%但图\ref{GACMAccuracy}和\ref{GACMCIFAR10Ln10}中恶意节点的比例仍对模型性能造成影响，这可能是因为在分配数据集时训练数据被分配到了恶意节点中，但监管节点查出恶意节点并阻止其继续参与联邦训练，使得可用于训练的数据减少，所以恶意节点比例越大，模型性能越小。
Figure \ref{GACMAccuracy} illustrates the relationship between the supervisory node's review of local models under our proposed MACM mechanism and model performance when varying proportions of malicious nodes are present among local training nodes (Ln), with the number of $Ln$ set to 10. Figure \ref{GACMCIFAR10Ln10} shows the impact of the supervisory node's review on model performance in the CIFAR10 dataset when the proportions of malicious nodes are \{0.1, 0.2\}. When the supervisory node does not review local models (S=0), model performance rapidly drops to between 0.1 and 0.2, with higher proportions of malicious nodes leading to lower performance. When the supervisory node reviews local models (S=1), model performance shows significant improvement compared to the case without reviews. Figure \ref{GACMFASHIONMNISTLn10} demonstrates the supervisory node's impact on model performance in the Fashion-MNIST dataset. Consistent with Figure \ref{GACMAccuracy}, the supervisory node's review of local models effectively enhances model performance. However, in both Figures \ref{GACMAccuracy} and \ref{GACMCIFAR10Ln10}, the proportion of malicious nodes still affects model performance. This may be due to training data being allocated to malicious nodes during dataset distribution. When the supervisory node identifies and excludes malicious nodes from further participation in federated training, the amount of data available for training decreases. Consequently, a higher proportion of malicious nodes results in lower model performance.

\subsubsection{Contribution-assessment Experiment}

\begin{figure*}[htbp]
	\centering
	%\begin{minted}[frame=single, linenos=true,breaklines=true]{json}
	\begin{subfigure}[b]{0.48\textwidth}
		\centering
		\captionsetup{justification=centering}
		\includegraphics[width=\textwidth]{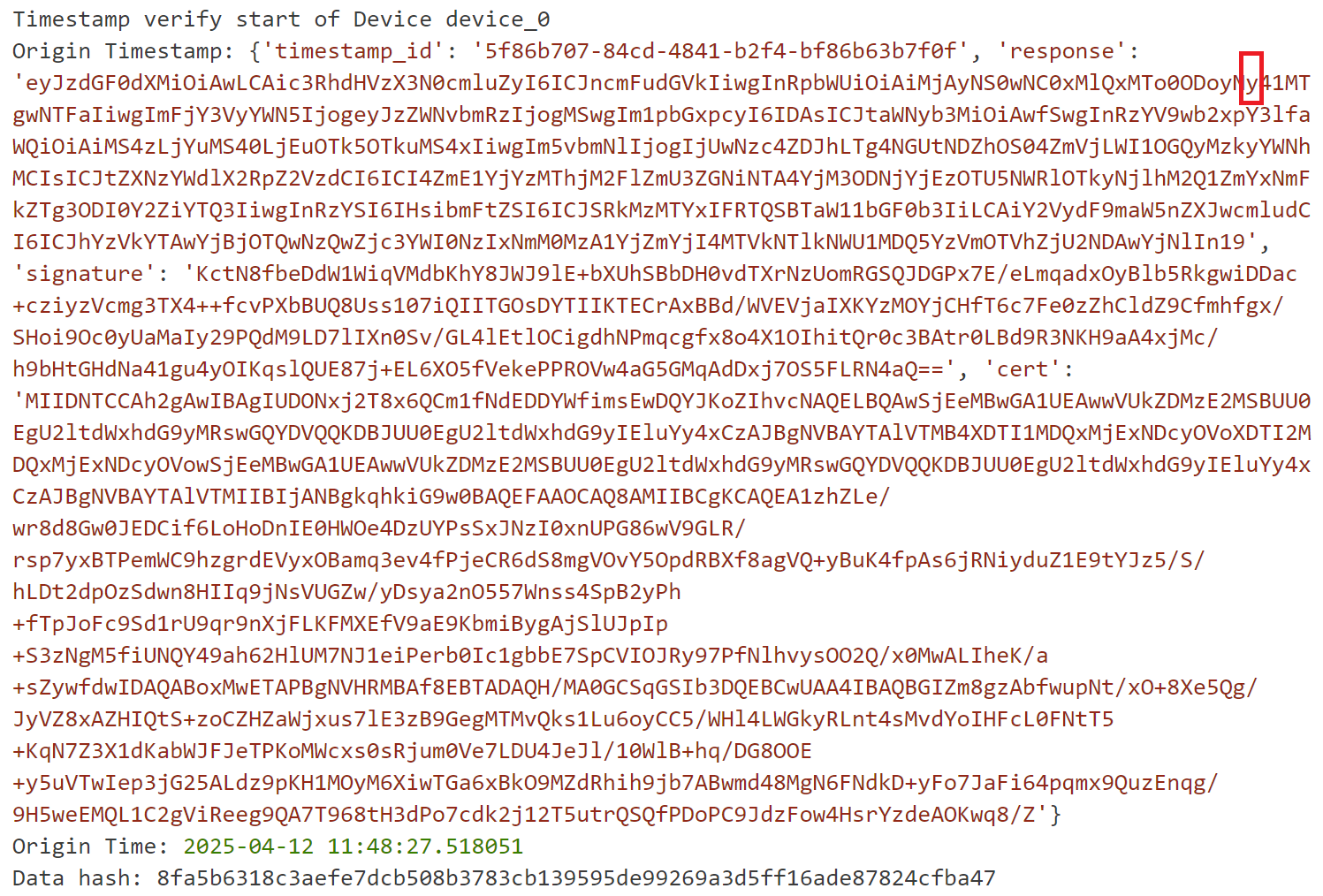}
		\caption{ Origin Timestamp}
		\label{OriginTimestamp}
	\end{subfigure} 
	\begin{subfigure}[b]{0.48\textwidth}
		\centering
		\captionsetup{justification=centering}
		\includegraphics[width=\textwidth]{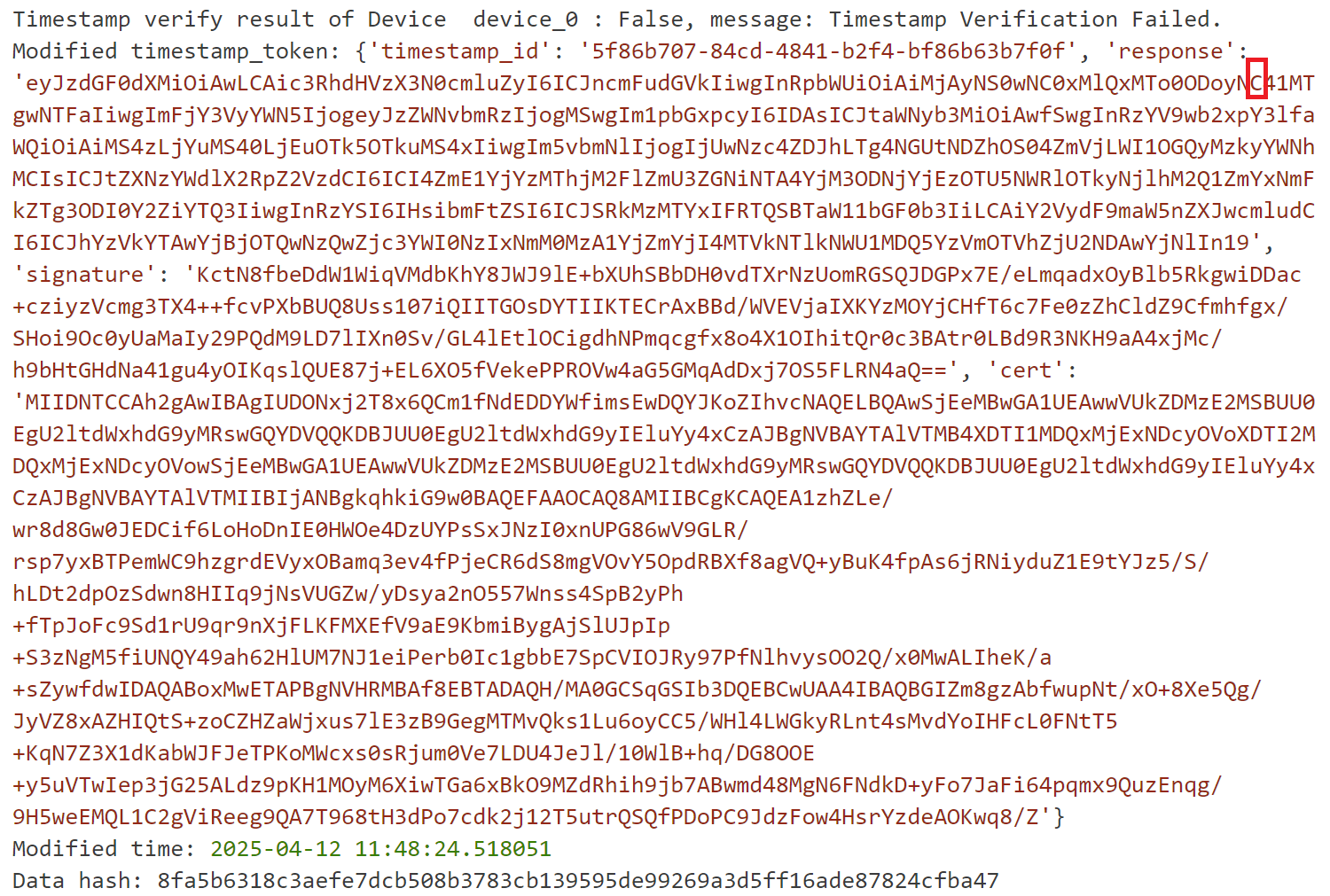}
		\caption{ Modified Timestamp}
		\label{ModifiedTimestamp}
	\end{subfigure}

	% \end{minted}
%\caption{Modified Timestamp Verification. (a):Origin trusted timestamp. (b). Modified trusted timestamp, 时间被篡改为真实时间的前3秒。恶意节点为了更多的奖励，试图篡改本地训练结束时间以减少训练的时间间隔，从而获得更多奖励。(b)中用红色标记出了可信时间戳篡改后与(a)不同的地方.  }
\caption{Modified Timestamp Verification. (a): Original trusted timestamp. (b): Modified trusted timestamp, altered to 3 seconds earlier than the actual time. To gain higher rewards, a malicious node attempts to tamper with the local training completion time to reduce the training time interval, thereby increasing its reward. In (b), the differences from (a) caused by the tampering of the trusted timestamp are highlighted in red.}
\label{MTV}
\end{figure*}

%图\ref{MTV}展示了在激励机制中可信时间戳的攻击和防御方式。\ref{OriginTimestamp}展示了恶意节点收到的可信时间戳，及其对应的时间。若恶意节点希望通过篡改可信时间戳，减少自己的训练时间，从而获得更多奖励时，它会选择提前自己的训练结束时间。我们模拟了恶意节点篡改可信时间戳的过程，通过解析时间戳篡改时间后发送给可信时间戳服务器，可信时间戳服务器会验证该时间戳是否被篡改。\ref{ModifiedTimestamp}中恶意节点篡改了收到的可信时间戳，将时间修改至3秒前，发送给可信时间戳服务器进行验证时被发现，可信时间戳返回验证结果为校验失败。Moreover, 对时间的篡改会导致时间戳发生变化，我们用红色的方框进行了标记。
Figure \ref{MTV} illustrates the attack and defense mechanisms involving trusted timestamps in the incentive mechanism. Figure \ref{OriginTimestamp} displays the trusted timestamp received by a malicious node and its corresponding time. If a malicious node attempts to tamper with the trusted timestamp to reduce its reported training time and thereby gain higher rewards, it may choose to advance its training completion time. We simulated the process of a malicious node tampering with the trusted timestamp. After modifying the timestamp, the node sends it to the trusted timestamp server, which verifies whether the timestamp has been altered. In Figure \ref{ModifiedTimestamp}, the malicious node tampers with the received trusted timestamp, changing the time to 3 seconds earlier. When this is sent to the trusted timestamp server for verification, the tampering is detected, and the server returns a verification failure result. Moreover, tampering with the time causes changes in the timestamp, which we have highlighted with a red box.

%
%
%图\ref{GACMStake}展示了我们提出的激励机制 IMTTI 的奖励方式，with number of $Ln$ is 10。。其中\ref{StakeGACMCIFAR10Ln101},\ref{StakeGACMCIFAR10Ln102},\ref{StakeGACMFASHIONMNISTLn101},\ref{StakeGACMFASHIONMNISTLn102}分别表示在CIFAR10，Fashion-MNIST数据集中恶意节点比例分别为\{0.1, 0.2\}的情况下，各个本地训练节点的stake变化。\ref{RewardGACMCIFAR10Ln101},\ref{RewardGACMCIFAR10Ln102}, \ref{RewardGACMFASHIONMNISTLn101}, \ref{RewardGACMFASHIONMNISTLn102}分别表示各个本地训练节点在每个轮次获得的reward变化。而\ref{LocalGACMCIFAR10Ln101}, \ref{LocalGACMCIFAR10Ln102}, \ref{LocalGACMFASHIONMNISTLn101}, \ref{LocalGACMFASHIONMNISTLn102}分别表示各个本地训练节点在每个轮次的训练时间。\ref{RewardGACMCIFAR10Ln101}-\ref{LocalGACMCIFAR10Ln101}, \ref{StakeGACMCIFAR10Ln102}-\ref{LocalGACMCIFAR10Ln102}, \ref{StakeGACMFASHIONMNISTLn101}-\ref{LocalGACMFASHIONMNISTLn101}, \ref{StakeGACMFASHIONMNISTLn102}-\ref{LocalGACMFASHIONMNISTLn102}分别为同一训练中的不同实验指标。
%对于恶意节点，其stake很快将下降至0，因为其被审查节点审查后禁止参与本地训练，如 $device_0$ in \ref{StakeGACMCIFAR10Ln101} and \ref{StakeGACMFASHIONMNISTLn101}, $device_0$ 和 $device_1$ in \ref{StakeGACMCIFAR10Ln102} and \ref{StakeGACMFASHIONMNISTLn102}. 由于每轮的总奖励是固定的，根据Equation \ref{Rrho} 每个训练节点获得的奖励和其训练时间成反比, 图\ref{GACMStake} 较好的印证了这一点。

\begin{figure*}[htbp]
	\centering
	
	\begin{subfigure}[b]{0.31\textwidth}
		\centering
		\captionsetup{justification=centering}
		\includegraphics[width=\textwidth]{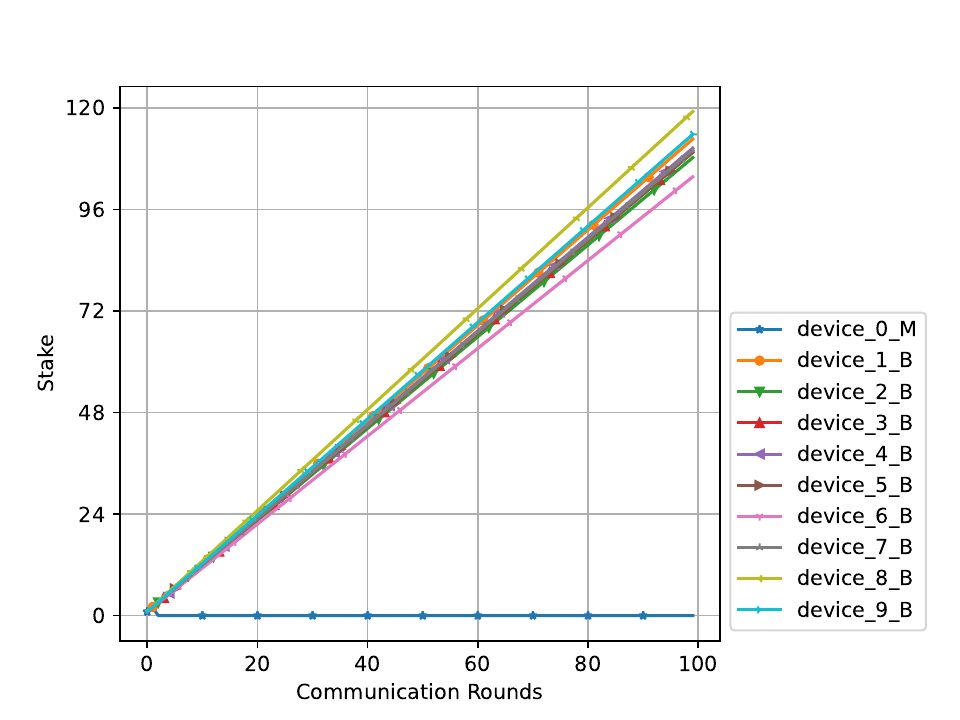}
		\caption{ Stake, CIFAR10, \\$Ln$=10, M =  0.1}
		\label{StakeGACMCIFAR10Ln101}
	\end{subfigure} 
	\begin{subfigure}[b]{0.31\textwidth}
		\centering
		\captionsetup{justification=centering}
		\includegraphics[width=\textwidth]{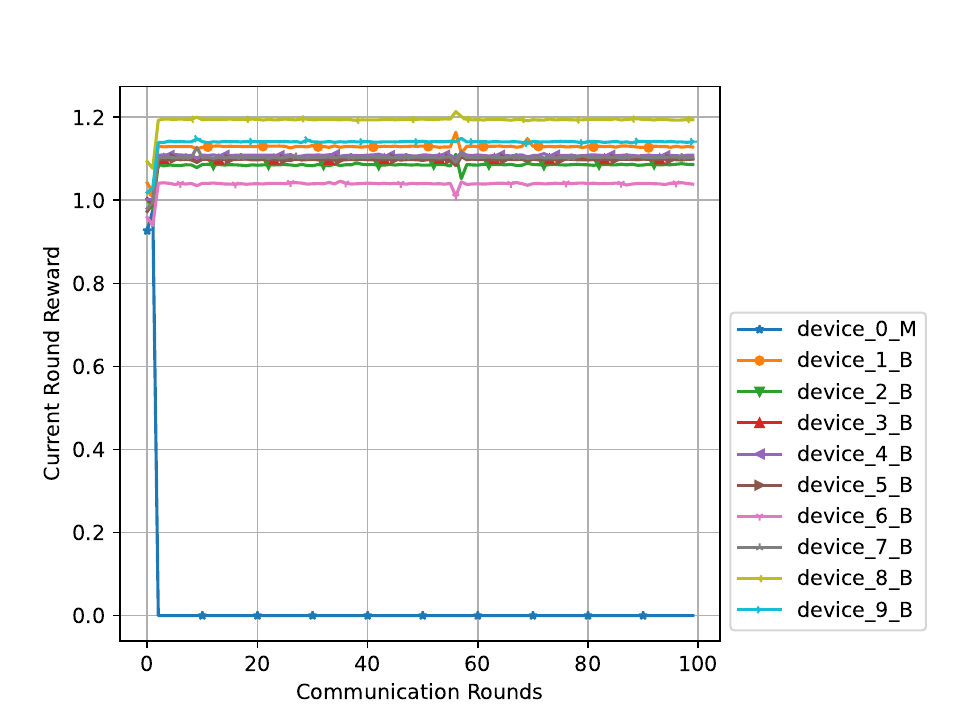}
		\caption{ Reward in One Round, CIFAR10, \\$Ln$=10, M =  0.1}
		\label{RewardGACMCIFAR10Ln101}
	\end{subfigure} 
	\begin{subfigure}[b]{0.31\textwidth}
		\centering
		\captionsetup{justification=centering}
		\includegraphics[width=\textwidth]{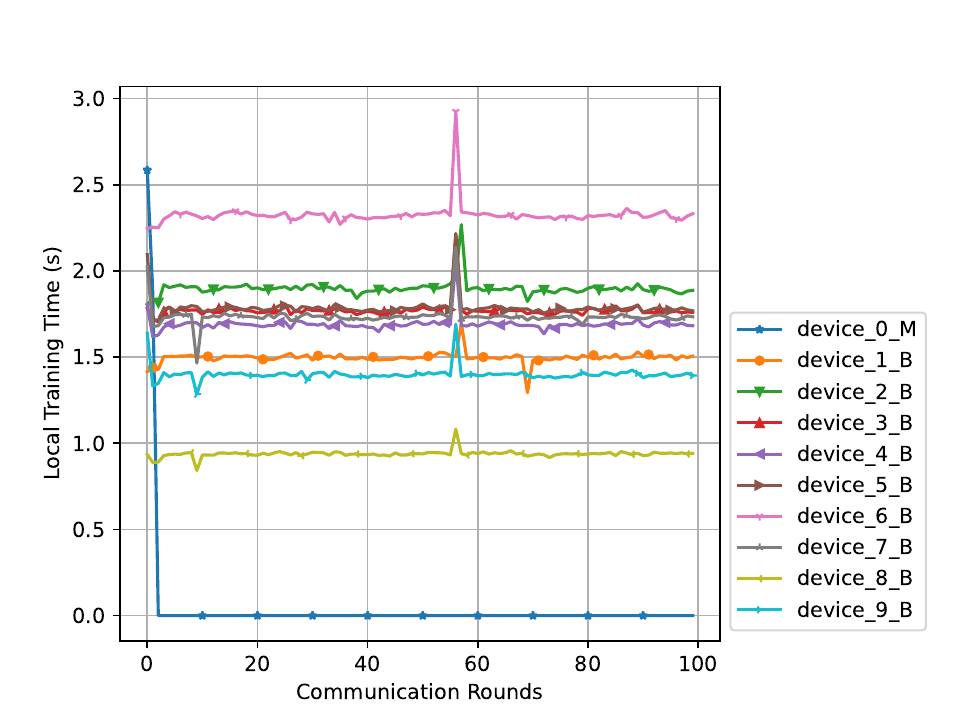}
		\caption{ Local Training Time, CIFAR10, \\$Ln$=10, M =  0.1}
		\label{LocalGACMCIFAR10Ln101}
	\end{subfigure} 
	
	\begin{subfigure}[b]{0.31\textwidth}
		\centering
		\captionsetup{justification=centering}
		\includegraphics[width=\textwidth]{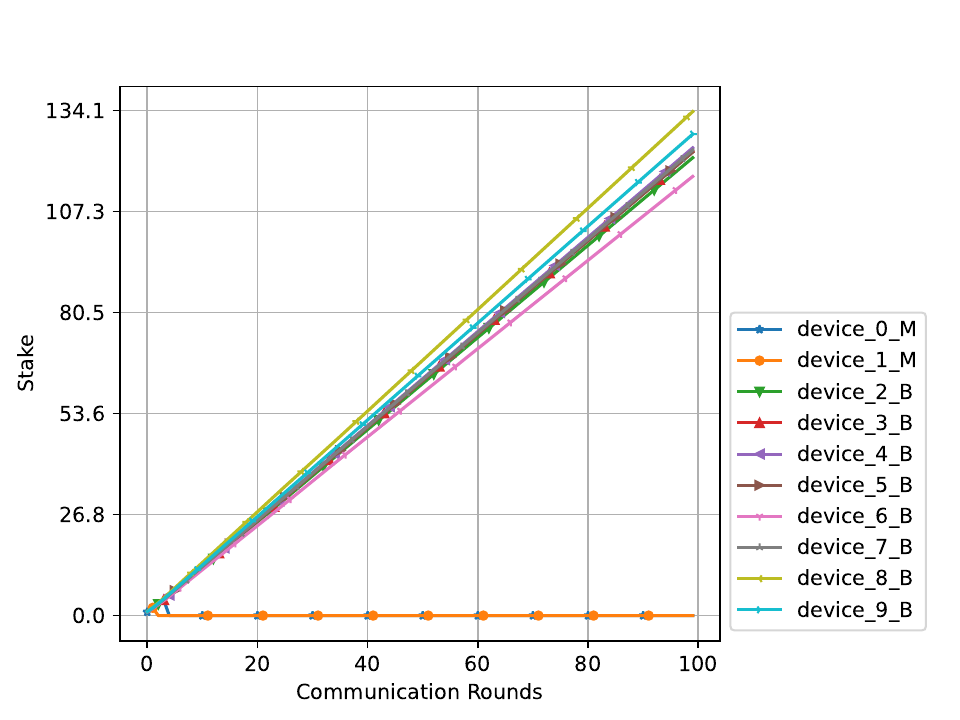}
		\caption{ Stake, CIFAR10, \\$Ln$=10, M =  0.2}
		\label{StakeGACMCIFAR10Ln102}
	\end{subfigure} 
	\begin{subfigure}[b]{0.31\textwidth}
		\centering
		\captionsetup{justification=centering}
		\includegraphics[width=\textwidth]{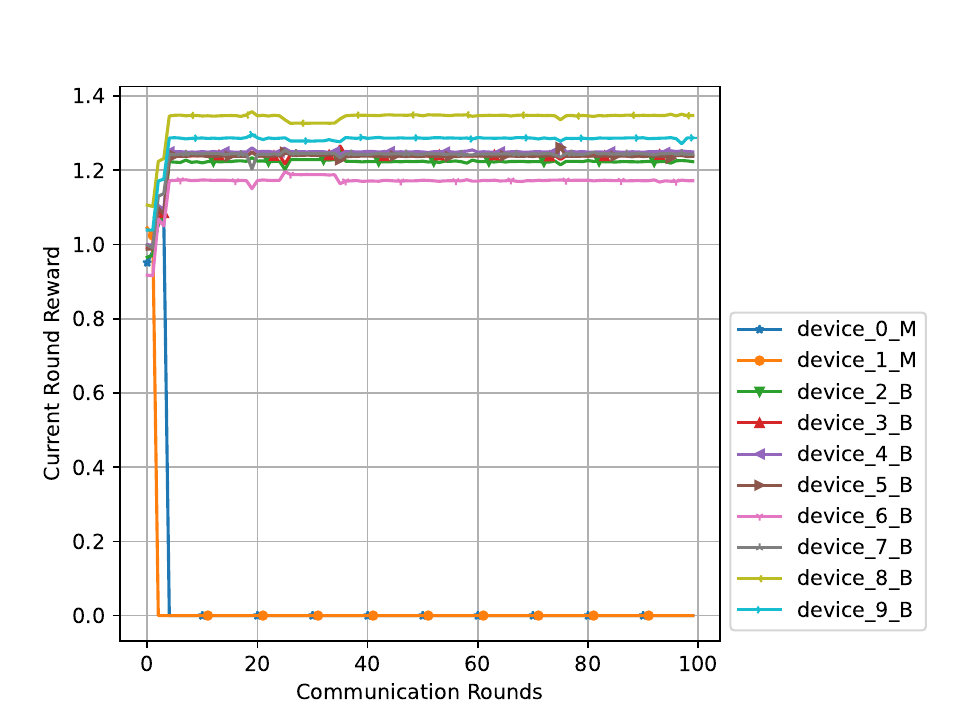}
		\caption{ Reward in One Round, CIFAR10, \\$Ln$=10, M =  0.2}
		\label{RewardGACMCIFAR10Ln102}
	\end{subfigure} 
	\begin{subfigure}[b]{0.31\textwidth}
		\centering
		\captionsetup{justification=centering}
		\includegraphics[width=\textwidth]{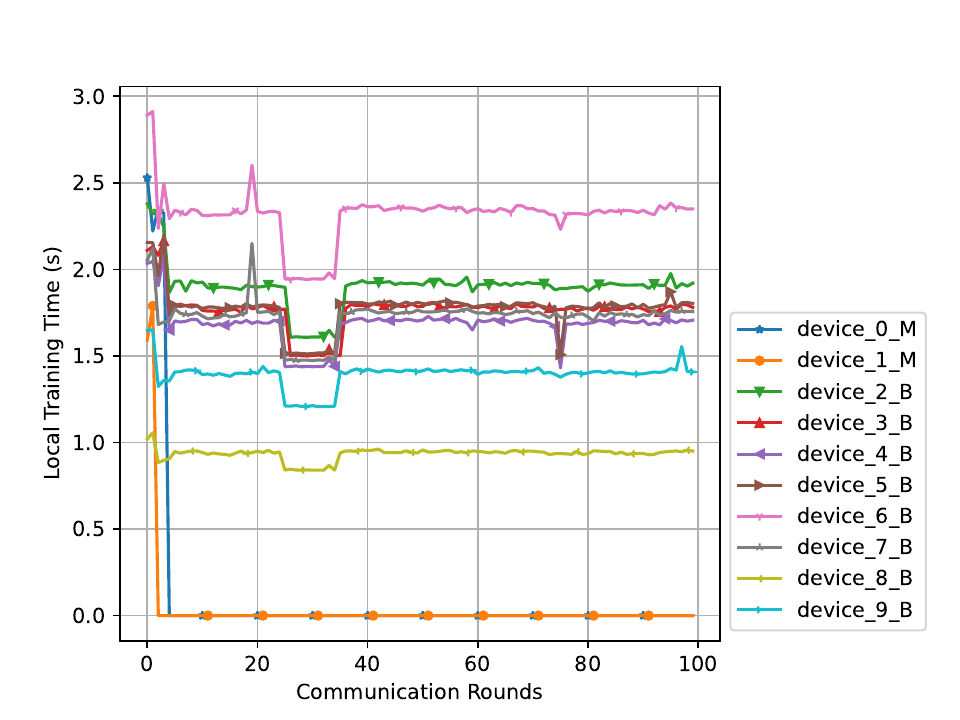}
		\caption{ Local Training Time, CIFAR10, \\$Ln$=10, M =  0.2}
		\label{LocalGACMCIFAR10Ln102}
	\end{subfigure} 
	
	\begin{subfigure}[b]{0.31\textwidth}
		\centering
		\captionsetup{justification=centering}
		\includegraphics[width=\textwidth]{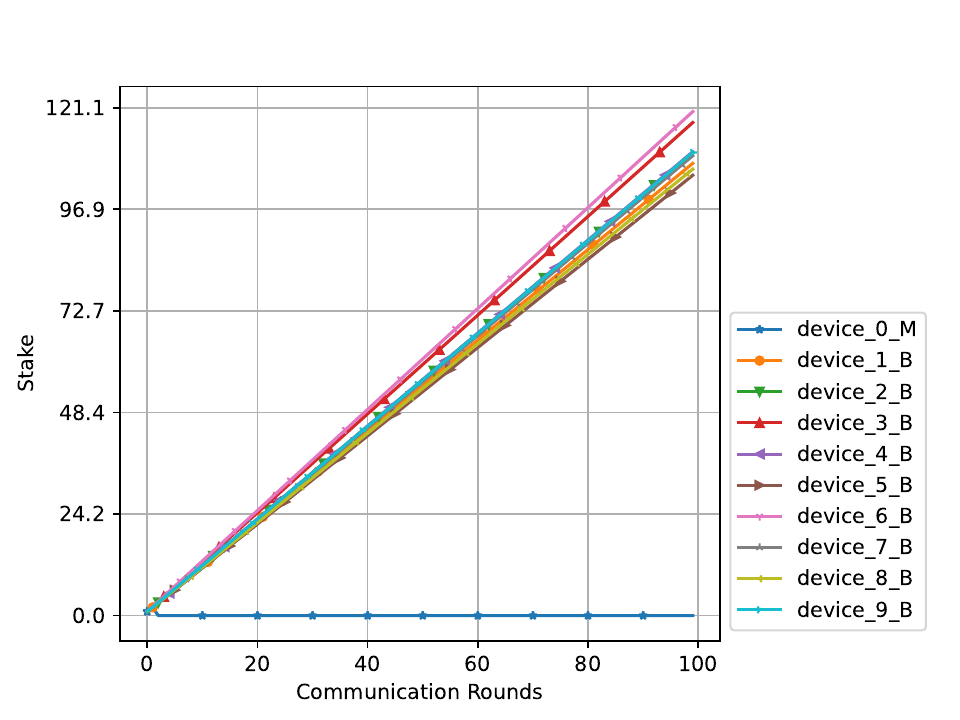}
		\caption{ Stake, Fashion-MNIST, \\$Ln$=10, M =  0.1}
		\label{StakeGACMFASHIONMNISTLn101}
	\end{subfigure} 
	\begin{subfigure}[b]{0.31\textwidth}
		\centering
		\captionsetup{justification=centering}
		\includegraphics[width=\textwidth]{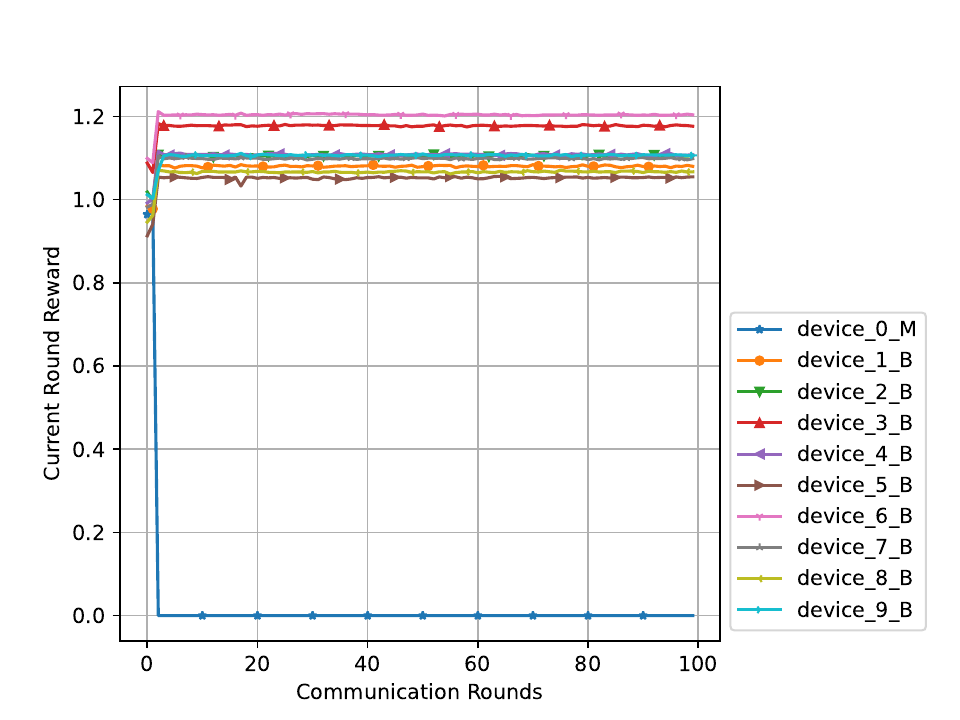}
		\caption{ Reward in One Round, Fashion-MNIST, $Ln$=10, M =  0.1}
		\label{RewardGACMFASHIONMNISTLn101}
	\end{subfigure} 
	\begin{subfigure}[b]{0.31\textwidth}
		\centering
		\captionsetup{justification=centering}
		\includegraphics[width=\textwidth]{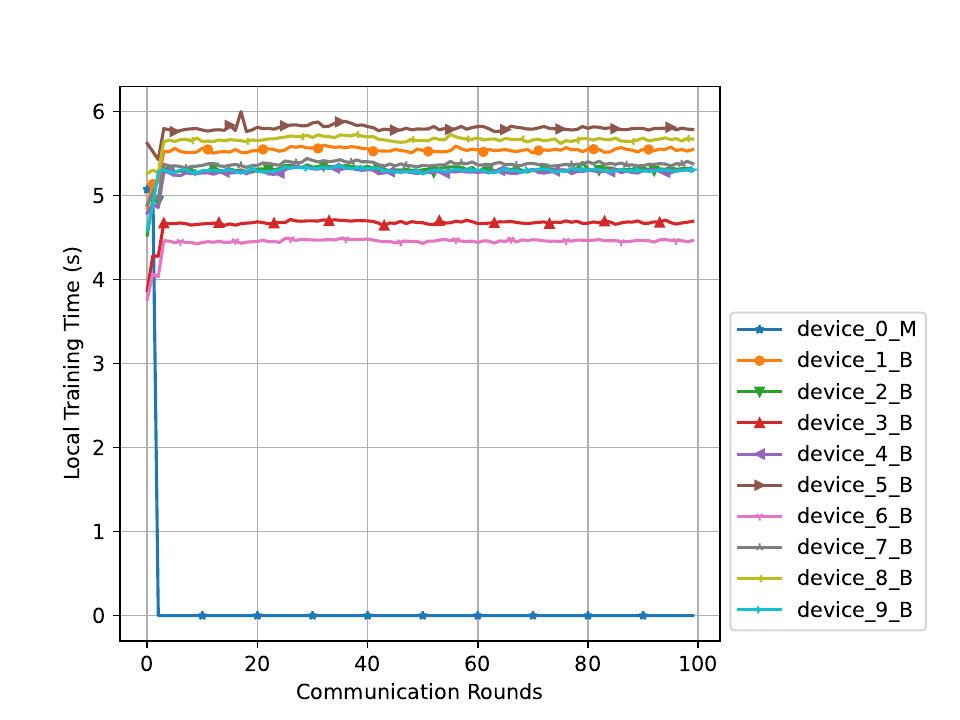}
		\caption{ Local Training Time, Fashion-MNIST, $Ln$=10, M =  0.1}
		\label{LocalGACMFASHIONMNISTLn101}
	\end{subfigure} 
	
	\begin{subfigure}[b]{0.31\textwidth}
		\centering
		\captionsetup{justification=centering}
		\includegraphics[width=\textwidth]{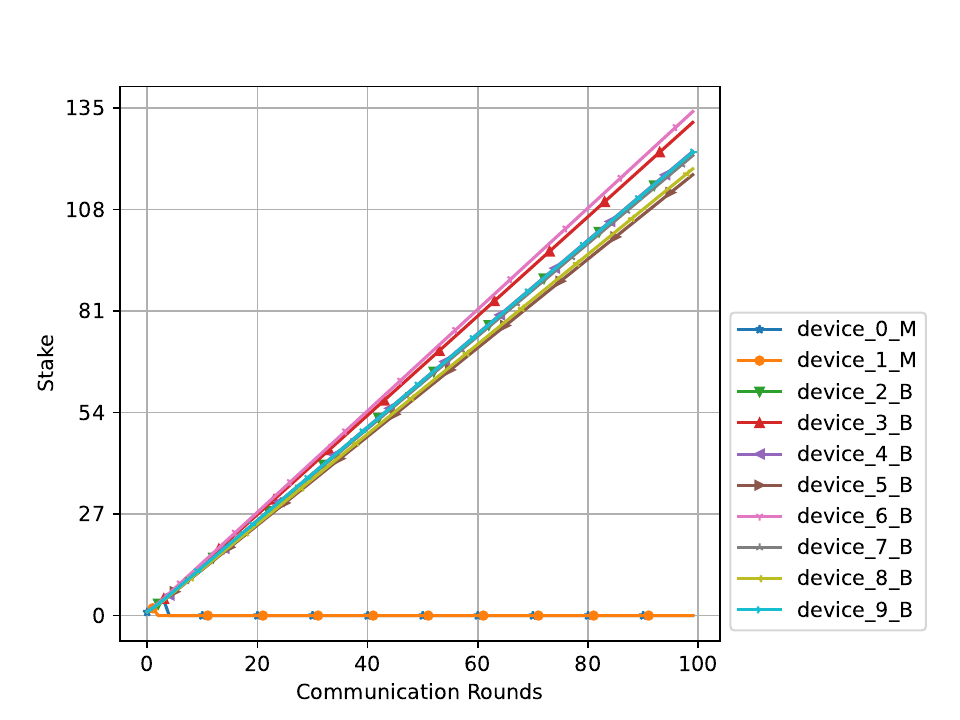}
		\caption{ Stake, Fashion-MNIST, \\ $Ln$=10, M =  0.2}
		\label{StakeGACMFASHIONMNISTLn102}
	\end{subfigure} 
	\begin{subfigure}[b]{0.31\textwidth}
		\centering
		\captionsetup{justification=centering}
		\includegraphics[width=\textwidth]{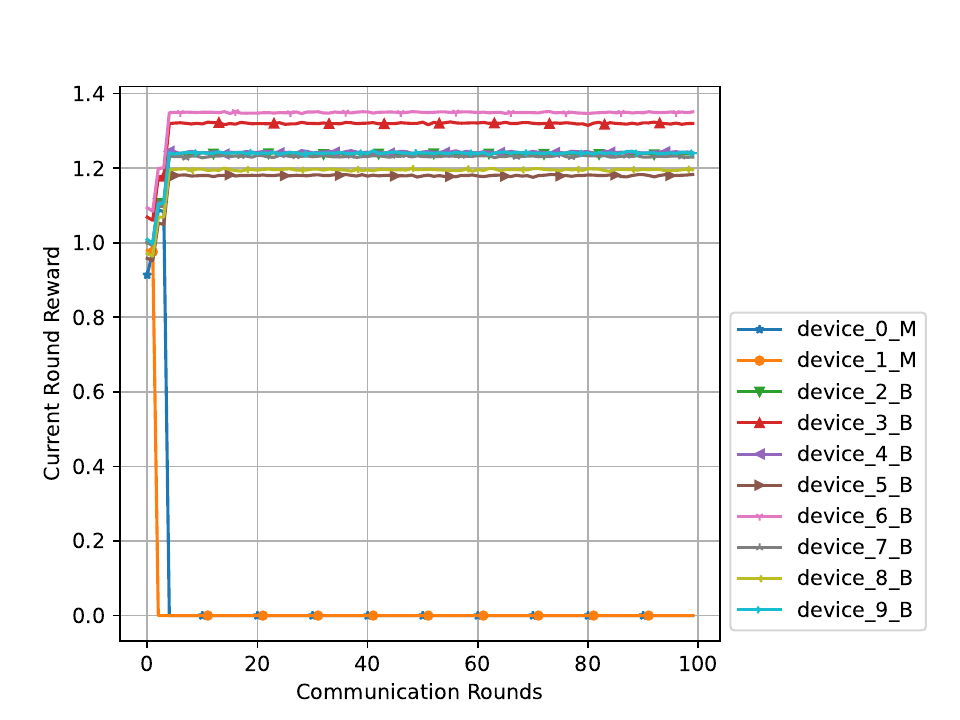}
		\caption{ Reward in One Round, Fashion-MNIST, $Ln$=10, M =  0.2}
		\label{RewardGACMFASHIONMNISTLn102}
	\end{subfigure} 
	\begin{subfigure}[b]{0.31\textwidth}
		\centering
		\captionsetup{justification=centering}
		\includegraphics[width=\textwidth]{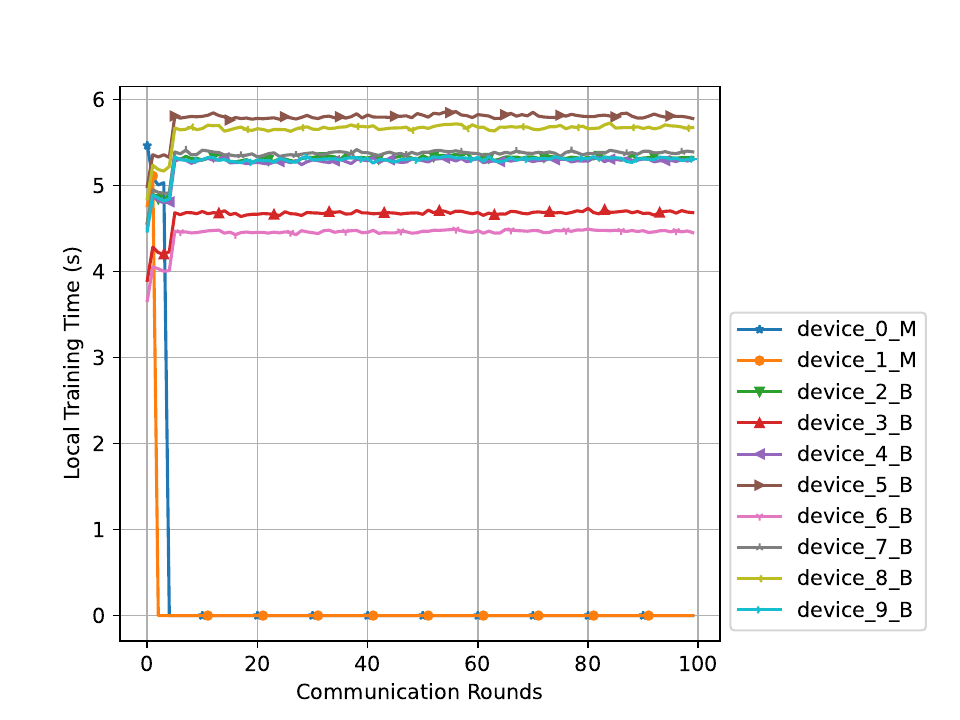}
		\caption{Local Training Time, Fashion-MNIST, $Ln$=10, M =  0.2}
		\label{LocalGACMFASHIONMNISTLn102}
	\end{subfigure}
	\caption{
		Different devices' stake, reward and local training time of HAM algorithm under Malicious Attack for CIFAR10 and Fashion-MNIST classification tasks. ``M" in the legend follows the $Ln$ idx means this node is malicious node, and ``B" in the legend means this node is benign node. ``m" in the caption represents the malicious nodes ratio in local training nodes. 
	}
	\label{GACMStake}
\end{figure*}

Figure \ref{GACMStake} illustrates the reward mechanism of our proposed IMTTI incentive mechanism, with the number of $Ln$ set to 10. Figures \ref{StakeGACMCIFAR10Ln101}, \ref{StakeGACMCIFAR10Ln102}, \ref{StakeGACMFASHIONMNISTLn101}, and \ref{StakeGACMFASHIONMNISTLn102} depict the stake changes of local training nodes in the CIFAR10 and Fashion-MNIST datasets when the proportions of malicious nodes are \{0.1, 0.2\}, respectively. Figures \ref{RewardGACMCIFAR10Ln101}, \ref{RewardGACMCIFAR10Ln102}, \ref{RewardGACMFASHIONMNISTLn101}, and \ref{RewardGACMFASHIONMNISTLn102} show the reward changes for each local training node across training rounds. Meanwhile, Figures \ref{LocalGACMCIFAR10Ln101}, \ref{LocalGACMCIFAR10Ln102}, \ref{LocalGACMFASHIONMNISTLn101}, and \ref{LocalGACMFASHIONMNISTLn102} illustrate the training time of each local training node per round. The pairs \ref{RewardGACMCIFAR10Ln101}-\ref{LocalGACMCIFAR10Ln101}, \ref{StakeGACMCIFAR10Ln102}-\ref{LocalGACMCIFAR10Ln102}, \ref{StakeGACMFASHIONMNISTLn101}-\ref{LocalGACMFASHIONMNISTLn101}, and \ref{StakeGACMFASHIONMNISTLn102}-\ref{LocalGACMFASHIONMNISTLn102} represent different experimental metrics within the same training setup.

For malicious nodes, their stake rapidly drops to 0 as they are identified by the supervisory node and barred from participating in local training. Examples include $device_0$ in \ref{StakeGACMCIFAR10Ln101} and \ref{StakeGACMFASHIONMNISTLn101}, and both $device_0$ and $device_1$ in \ref{StakeGACMCIFAR10Ln102} and \ref{StakeGACMFASHIONMNISTLn102}. Since the total reward per round is fixed, according to Equation \ref{Rrho}, the reward each training node receives is inversely proportional to its training time. Figure \ref{GACMStake} effectively validates this relationship.

%\subsection{综合性能评估}
%\subsubsection{端到端系统性能}
%实验目的: 评估FLSSM整体系统的性能
%
%对比方法:
%传统联邦学习/
%基于同态加密的联邦学习/
%完整FLSSM系统
%
%评估指标:
%模型准确率/
%训练时间/
%系统开销/
%安全性能
%
%
%\subsubsection{消融实验}
%实验目的: 验证FLSSM各组件的贡献度
%
%实验配置:
%完整FLSSM/
%FLSSM无HAM/
%FLSSM无MACM/
%FLSSM无IMTTI
%
%评估指标: 与完整系统性能的对比

%\subsection{ Ablation Experiment}
%\subsubsection{attack-tracing}
%3个数据集
%受到恶意攻击时，对恶意攻击的检测和溯源
%实现2种不同的恶意攻击（噪声攻击、xx攻击），设置安全聚合算法（KMeans） 进行检测
%将报错信息贴出来
%\subsubsection{contribution-assessment}
%3个数据集
%节点的训练时间、token奖励 画图
%可信时间戳 格式 代码贴出来

\section{ Conclusion}
\label{Conclusion}
%This artical proposed the FLSSM for encrypted federated learning model to address the problems of "computation-efficiency","attack-tracing" and "contribution-assessment". 首先，我们采用边缘聚合节点，提高加密模型聚合效率，从而解决"computation-efficiency"问题。其次，我们提出模型访问控制机制，由第三方可信节点对加密后的模型进行审查，以检测恶意攻击，从而解决"attack-tracing"问题。最后，我们提出一种激励机制，通过可信时间戳服务器实现本地模型训练时间的可信性，在模型加密的前提下对本地训练节点的贡献进行公平评估，从而解决同态加密环境下的"contribution-assessment"问题。两个现实世界的公开数据集验证了我们想法的可行性。结果表明，我们的模型不仅可以提高聚合效率，追溯恶意节点，还能有效评估节点所做贡献。
This article proposes FLSSM, a framework for encrypted federated learning models, to tackle the challenges of ``computation-efficiency," ``attack-tracing," and ``contribution-assessment." First, we employ edge aggregation nodes to enhance the aggregation efficiency of encrypted models, addressing the ``computation-efficiency" issue. Second, we introduce a model access control mechanism, where a trusted third-party node reviews encrypted models to detect malicious attacks, resolving the ``attack-tracing" problem. Finally, we propose an incentive mechanism that leverages a trusted timestamp server to ensure the reliability of local model training times, enabling fair contribution assessment of local training nodes under homomorphic encryption, thus addressing the ``contribution-assessment" challenge. Experiments on two real-world public datasets validate the feasibility of our approach. The results demonstrate that our model not only improves aggregation efficiency and enables tracing of malicious nodes but also effectively evaluates node contributions.
\section{ ACKNOWLEDGMENTS}

This work was supported by the National Natural Science Foundation of China under Grant No. 62272024.

%\section{Introduction}
%\IEEEPARstart{T}{his} file is intended to serve as a ``sample article file''

%\begin{thebibliography}{1}
%\bibliographystyle{IEEEtran,reference}
%\end{thebibliography}
\bibliographystyle{IEEEtran}
\bibliography{reference.bib}

\newpage

% your own custom macro containing the $\backslash${\tt{includegraphics}} command to make things
%\bf{If you include a photo:}\vspace{-33pt}
%\begin{IEEEbiography}[{\includegraphics[width=1in,height=1.25in,clip,keepaspectratio]{fig1}}]{Michael Shell}
%Use $\backslash${\tt{begin\{IEEEbiography\}}} and then for the 1st argument use $\backslash${\tt{includegraphics}} to declare and link the author photo.
%Use the author name as the 3rd argument followed by the biography text.
%\end{IEEEbiography}

\vfill

\end{document}